\documentclass[%
aip,
 amsmath,amssymb,
 reprint,
]{revtex4-1}

\usepackage{graphicx}
\usepackage{xcolor}
\usepackage{dcolumn}
\usepackage{bm}
\usepackage{hyperref}
\usepackage[utf8]{inputenc}
\usepackage[T1]{fontenc}
\usepackage{mathptmx}
\usepackage{amsmath}
\usepackage[switch]{lineno} 
\usepackage{ulem}
\usepackage[version=3]{mhchem} 

\usepackage{xcolor}

\newcommand*\subt[1]{_{\textnormal{#1}}}
\newcommand*\supt[1]{^{\textnormal{#1}}}

\begin{document}
\affiliation{Department of Physics, King's College London, London, WC2R 2LS, UK}
\affiliation{College of Engineering, Swansea University, Bay Campus, Fabian Way, Swansea, SA1 8EB, UK}
\affiliation{Present address: International School for Advanced Studies, Via Bonomea, 265, 34136, Trieste, IT}
\affiliation{Present address: Laboratory of Nanochemistry, Institute of Chemistry and Chemical Engineering, Ecole Polytechnique Fédérale de Lausanne, Lausanne, CH}
\affiliation{Present address: Department of Physics, Aristotle University of Thessaloniki, Thessaloniki GR-54124, GR}
\affiliation{Present address: College of Engineering, Swansea University, Bay Campus, Fabian Way, Swansea, SA1 8EB, UK}
\affiliation{Present address: DIPC, Paseo Manuel de Lardizabal, 20018 San Sebastian, Spain}
\affiliation{email: czeni@sissa.it}

\author{Claudio Zeni}
\affiliation{Department of Physics, King's College London, London, WC2R 2LS, UK}
\affiliation{Present address: International School for Advanced Studies, Via Bonomea, 265, 34136, Trieste, IT}
\affiliation{email: czeni@sissa.it}
\author{Kevin Rossi}
\affiliation
{Department of Physics, King's College London, London, WC2R 2LS, UK}
\affiliation{Present address: Laboratory of Nanochemistry, Institute of Chemistry and Chemical Engineering, Ecole Polytechnique Fédérale de Lausanne, Lausanne, CH}
\author{Theodore Pavloudis}
\affiliation{College of Engineering, Swansea University, Bay Campus, Fabian Way, Swansea, SA1 8EB, UK}
\affiliation{Present address: Department of Physics, Aristotle University of Thessaloniki, Thessaloniki GR-54124, GR}
\author{Joseph Kioseoglou}
\affiliation{Present address: Department of Physics, Aristotle University of Thessaloniki, Thessaloniki GR-54124, GR}
\author{Stefano de Gironcoli}
\affiliation{Present address: International School for Advanced Studies, Via Bonomea, 265, 34136, Trieste, IT}
\author{Richard E. Palmer}
\affiliation{Present address: College of Engineering, Swansea University, Bay Campus, Fabian Way, Swansea, SA1 8EB, UK}
\author{Francesca Baletto}
\affiliation
{Department of Physics, King's College London, London, WC2R 2LS, UK}
\affiliation{Present address: DIPC, Paseo Manuel de Lardizabal, 20018 San Sebastian, Spain}

\title{Data-driven simulation and characterisation of gold nanoparticle melting}








\begin{abstract}
\section*{Abstract}
The simulation and analysis of the thermal stability of nanoparticles, a stepping stone towards their application in technological devices, require fast and accurate force fields, in conjunction with effective characterisation methods.
In this work, we develop efficient, transferable, and interpretable machine learning force fields for gold nanoparticles based on data gathered from Density Functional Theory calculations.
We use them to investigate the thermodynamic stability of gold nanoparticles of different sizes (1 to 6 nm), containing up to 6266 atoms, concerning a solid-liquid phase change through molecular dynamics simulations.
We predict nanoparticle melting temperatures in good agreement with available experimental data.
Furthermore, we characterize the solid-liquid phase change mechanism employing an unsupervised learning scheme to categorize local atomic environments.
We thus provide a data-driven definition of liquid atomic arrangements in the inner and surface regions of a nanoparticle and employ it to show that melting initiates at the outer layers.
\end{abstract}

\maketitle

\section*{Introduction}
Gold (Au) nanoparticles (NPs) find widespread application in many technological areas, such as in optics,\cite{Amendola2017,Jauffred2019} nanomedicine,\cite{Dreaden2012,Carnovale2019} and catalysis \cite{Ha2018,Zhang2020,Carter2017,Mitsudome2013,Zhao2019}.
As all chemo-physical properties of Au NPs depend on their shape, the analysis of their structural stability has attracted a lot of attention in the past years.
A deep understanding of the liquid-solid phase change mechanisms of Au NP, also accounting for surface rearrangements, may, in particular, be crucial for catalytic applications, where the reaction conditions often demand the nanocatalysts to work at high temperatures while preserving their size and shape.

Numerical simulations can, in principle, offer a platform to investigate and characterize phase change mechanisms of NPs at an atomistic level.
However, two long-standing challenges must be overcome to improve the numerical predictions of NPs' thermal stability.
The first concerns the difficulty of defining an unbiased characterisation of the phase change mechanism.
Indeed the identification of order parameters to characterise solid-liquid phase changes at the nanoscale is an active topic of debate with a long tradition.\cite{Gilvarry1956, Honeycutt1987, Lechner2008, Delgado-Callico2020}
Widely used methods often rely on chemical-intuition and heuristic approaches, and can therefore lead to descriptive order parameters which are neither fully general nor robust to parameter tuning.
For example, changes in the first neighbour distance distribution affect the definition of coordination number too drastically, \cite{Li2008a, Fukuya2020} and little research has been carried out on the characterization of local atomic environments peculiar of NP's surface atoms.

The second challenge is related to the development of accurate and fast interparticle potentials, which reproduce the complexity of the NPs' energy landscape. 
In so far, atomistic modelling methods have offered a strict trade-off between computational speed and accuracy.
While simulations based on electronic structure methods, such as density functional theory (DFT), provide quantitative accuracy, their computational cost severely limits the capabilities to generate dynamical trajectories of large systems and for long times. 
On the contrary, large systems and long simulation timescales are easily accessible when employing semi-empirical potentials.
Nevertheless, such methods do not necessarily provide a quantitative insight on the chemistry of NPs' phase changes \cite{Ellaby2018} because their analytical functional form limits their predictive power and flexibility.
Furthermore, these potentials are often fitted to bulk properties, which poses an additional limit to their accuracy when simulating nanoscale systems \cite{Baletto2002}.

In this work, we tackle these two challenges by adopting data-driven methods to generate an accurate and efficient description of interatomic potentials, and by developing an automated routine that classifies the atomic environments observed during Au NPs' phase change.
To obtain long, i.e., hundreds of ns, and accurate trajectories during melting of Au NPs of variable sizes, we develop a set of machine learning force fields (ML-FFs)\cite{Rossi2020, Cheng2019, Cooper2020, Deringer2019, Behler2017, Zeni2019, lot2020, xie2021bayesian} using the innovative framework of mapped Gaussian processes \cite{Glielmo2018, Zeni2018, vandermause2020fly}.
ML-FFs can approximate the force-energy predictions yielded by the reference DFT method they are trained upon while being many orders of magnitude faster to compute. 
Here, we train ML-FFs on LDA-DFT data, rPBE-DFT data, and contrast our results with experimental results, and with predictions found using a semi-empirical interatomic potential.
To characterise the melting kinetics, we adopt an unsupervised machine learning clustering scheme which discriminates in an automatic fashion locally liquid from locally solid environments, surface from inner environments, and high-coordination from low-coordination surface environments. 
We then obtain a route to estimate the NPs melting temperature by monitoring the relative population of liquid atoms in the nanoparticle, and the melting mechanism by recording the spatial distribution of locally liquid environments as a function of temperature.
We employ these data-driven tools to study the melting of Au NPs with diameters between 1 and 6~nm, and various initial geometrical shapes. 
We univocally show that melting initiates in the outermost layer of Au NPs first, and occurs in the NPs' core second.

\section*{Results}
\label{results}
\subsection*{Machine Learning Force Fields}
\label{subsec:mlff_results}
To construct a training database, we extract 7 random de-correlated frames from ab initio molecular dynamics trajectories where an Au NP containing 309 atoms ($\sim$2~nm of diameter) with an initial face centred cubic (FCC) morphology undergoes melting.
We calculate, for each frame, forces and energies at DFT LDA and DFT GGA-rPBE levels, and utilize a 2+3-body mappable Gaussian process regression framework \cite{Glielmo2018, Zeni2018} to fit two ML-FFs, one for each DFT method; further detail is provided in the Methods and in the Supplementary Methods.
When training on 2100 local atomic environments,
our ML-FFs incur in a mean absolute error (MAE) on the force components of 0.09~$\pm$~0.04~eV/$\text{\AA}$ (LDA ML-FF) and 0.07~$\pm$~0.03~eV/$\text{\AA}$ (rPBE ML-FF), and in a MAE on the atomic energy differences of 2.65 $\pm$ 2.02~meV/atom (LDA ML-FF) and 1.98~$\pm$~1.76~meV/atom (rPBE ML-FF), on validation sets disjointed from the training sets.
The reported accuracy is comparable to the ones quoted in previous studies\cite{Jindal2017,Jindal2018,Jindal2020,Thorn2019,Loeffler2020,Li2018,Chiriki2017}, and is deemed satisfactory.
This training dataset, albeit small, contains a heterogeneous set of local atomic environments, as shown in Supplementary Figure~1; and we, therefore, consider it to be representative for Au NPs in the size range of interest.

We test the accuracy of the two ML-FFs on a more complex dataset, which encompasses NPs' architectures of different sizes and shapes (see Supplementary Methods).
Supplementary Table~1 reports the mean absolute errors (MAEs) on force components and atomic energies incurred by each of the ML-FFs developed on these validation datasets.
The MAEs on force components are again consistently around 0.1~eV/$\text{\AA}$, and the MAEs on atomic energy differences are consistently lower than 10~meV/atom.
The ML-FFs are, therefore, considered accurate enough and, more importantly, transferable across different NPs' sizes and shapes.
This holds regardless of the DFT level of theory used to train the ML-FF (GGA PBE and LDA) and its implementation (VASP projector augmented-wave and CP2K Gaussian plane wave).

When validating the ML-FFs against the experimental bulk cohesive energy (Supplementary Figure~4), we observe that LDA (rPBE) based ML-FF overestimates (underestimates) this quantity.
We then adopt a parametric mixing of the two ML-FFs (see also the Methods section) and generate a third ML-FF, labelled hybrid, which, by construction, has cohesive energy in the bulk phase that matches the experimental one.
The 2- and 3-body FFs forming the three ML-FFs present some noticeable differences; in Supplementary Figure~5 we show how the LDA ML-FF is more bound and stiffer than the rPBE ML-FF, and how the hybrid ML-FF has, as expected, a shape that is in-between the one of the other two ML-FFs.

\subsection*{Phase Change Characterization}
\label{subsec:phase_change_results}

Following the successful training and validation of our ML-FFs, we employ them to study the size-dependent melting temperature of Au NPs. 
We consider NPs whose diameter ranges from 1 to 6~nm, corresponding to NPs containing 147, 309, 561, 923, 2869, and 6266 atoms.
We sample the NPs' evolution in a temperature range between 400K and 1600K when subject to a heating rate of 20~K/ns.
We also test 5~K/ns and 10~K/ns heating rates and do not observe significant changes in the melting temperature estimate (see Supplementary Figure~6).
For each NP size and ML-FF, we simulate Au NPs for a total of 2.4~ms, a time-length not accessible to electronic structure calculations, even for the largest state-of-the-art computational facilities.
We refer the interested reader to the Methods section for further details on the numerical setup used to perform the simulations.

\begin{figure*}
    \centering
    \includegraphics[width=15cm]{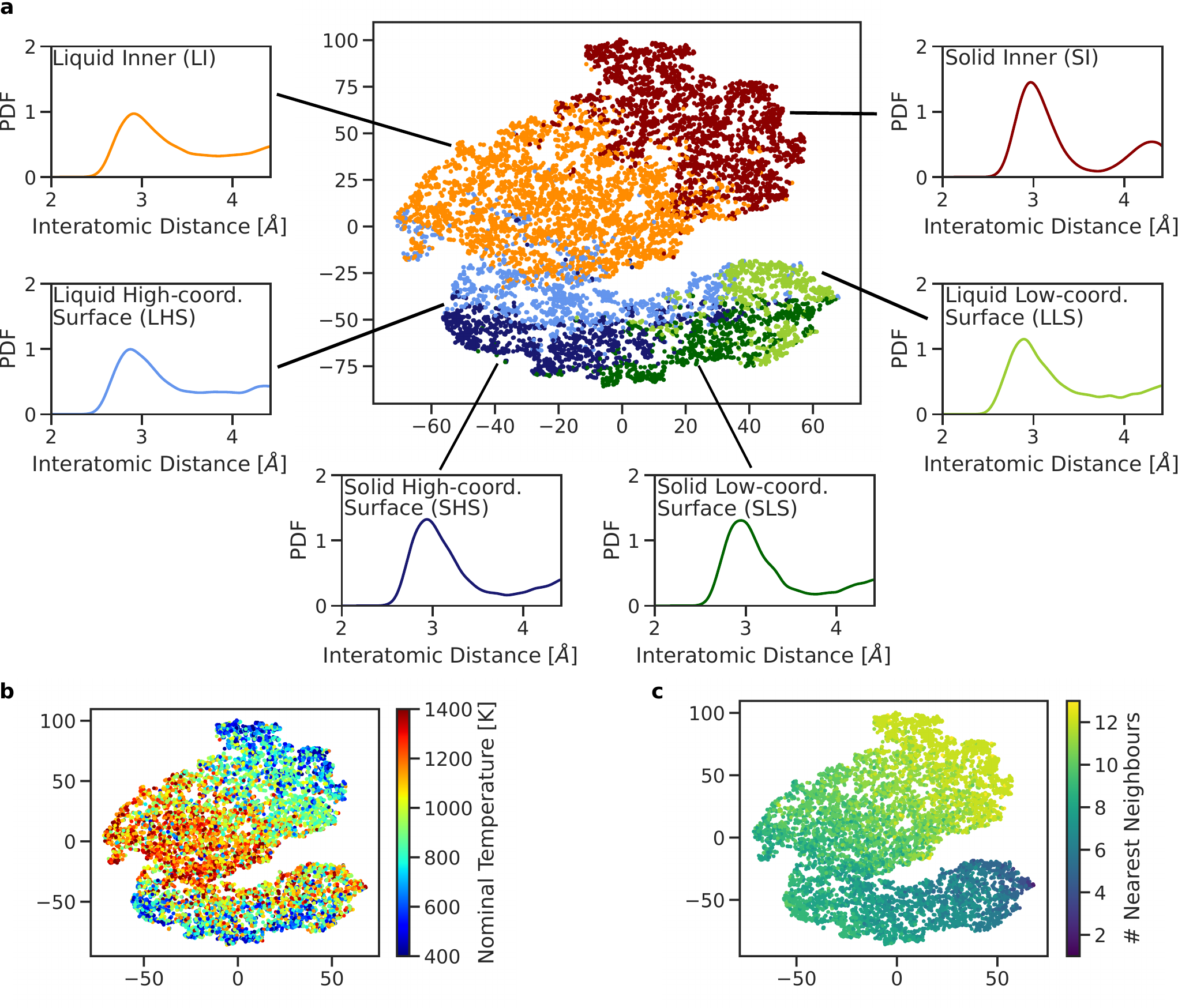}
    \caption{
    \textbf{Features of the six classes of local atoms environments identified through clustering.}%
    Visualization of the hierarchical k-means clustering results for MD simulations of Au nanoparticles with 147, 309, 561, 923, 2869 and 6266 atoms, carried out using the ML-FF trained on rPBE-DFT data.
    Panel a: 1st and 2nd component (x- and y-axis) of the t-sne projection of the atomic expansion coefficients of 10$^4$ local atomic environments randomly sampled from melting MD simulations. 
    The colours label the six classes assigned by the hierarchical k-means clustering algorithm, as defined in the main text.
    The normalized average pair-distance distribution function (PDF) belonging to each class is shown and coloured accordingly. 
    Panels b and c: same t-sne projection as in panel a). 
    In panel b, the colours indicate the nominal simulation temperature at which the local atomic environment was taken from, in panel c, the number of nearest neighbours (\#NN) computed using a cut-off of 3.6~$\text{\AA}$.
    }
    \label{fig:maps}
\end{figure*}

To characterize the solid-liquid phase transition, while distinguishing between surface and inner melting, we adopt an unsupervised machine learning approach that hinges on a small database of configurations randomly extracted from the phase change trajectories we simulated and a local atomic density representation.
In particular, we employ a modified version \cite{Zeni2021} of the 3-body local atomic cluster expansion descriptor \cite{Drautz2019} to associate a 40-dimensional set of features to each atom.
We then exploit a hierarchical k-means clustering scheme to isolate six classes of local atomic environments (see also the Methods and Supplementary Methods).
This labels the local atomic environments as being in solid or liquid phase, and as belonging to the inner,
high-coordination surface, and low-coordination surface motifs.
As illustrated in panels b and c of Figure~\ref{fig:maps} and Supplementary Figures 6 and 13, both the number of nearest neighbours (\#NN) within a predefined cut-off and the nominal MD simulation temperature at which these are sampled, correlate with the labels assigned by the clustering algorithm.

The six local atomic environment classes showcased in Figure~\ref{fig:maps} and Supplementary Figures 8 and 9, can be characterised from the number of  neighbours within a given cut-off (here taken as 3.50~$\text{\AA}$ for the LDA ML-FF, 3.75~$\text{\AA}$ for the rPBE ML-FF, and 3.60~$\text{\AA}$ for the hybrid ML-FF) they display, and from features in their pair-distance distribution function (PDF).
In detail:
\begin{itemize}
    \item Solid inner (SI) atoms have 12 NN within the chosen cut-off, and their PDF displays a well-defined peak at a second NN distance consistent with the one of bulk FCC Au. 
    SI local atomic environments comprise FCC-like motifs, as well as motifs with 5-fold- or icosahedral-symmetry.
    \item Liquid inner (LI) atoms have, on average, 11 NNs.
    The PDF for this class of local atomic environment presents the first peak at distances lower than the one for bulk lattice and lacks a pronounced second peak in correspondence to the bulk lattice one.
    \item Solid High-coordination Surface (SHS) atoms present, on average, 8 NNs, and peaks its PDF in correspondence to the second nearest-neighbours (lattice bulk).
    \item Liquid High-coordination Surface (LHS) atoms also have 8 NNs on average, yet the PDF lacks a peak at the bulk lattice.
    \item Solid Low-coordination Surface (SLS) atoms find an average of 6.9 atoms at a distance consistent with the bulk NNs distance.
    \item Liquid Low-coordination Surface (LLS) atoms have, on average, 6.0 atoms at a distance lower than the bulk nearest-neighbours distance; furthermore, the PDF does not display any peak for the second NNs.
\end{itemize}

This unsupervised approach enables an original and data-driven definition of liquid atomic arrangements in the inner part of the NPs and at the surface. 
Local atomic environments in a liquid phase are all characterized by the absence of a peak of their PDF in correspondence to their bulk lattice distance (i.e., the one predicted by the reference interatomic potential).
This observation holds regardless of whether they lie in the inner part or at the surface of the NP, and their coordination.
Furthermore, this result confirms and rationalises the universal signature of melting for the whole NP we recently proposed \cite{Delgado-Callico2020}.

Having discriminated in automated fashion atoms in liquid and solid environments, also as a function of their spatial location in the NP, we draw novel definitions to determine melting phase changes in the nanoparticle.
To this end we monitor the time evolution of the occurrence of liquid environments, their rate of variation, and at which temperature their relative population increases above the 0.4 of the total, also as a function of their distance from the center of mass of the NP.
In the following, we refer to the melting temperature of the NP ($T\subt{melt}\supt{NP}$) as the temperature at which the number of inner atoms that are identified as liquid (\#LI) by the clustering algorithm displays the maximum positive derivative.
This melting temperature estimation method yields equivalent results w.r.t. other well-established algorithms to calculate the melting temperature, such as the caloric curve maximum derivative and the heat capacity peak (see Supplementary Figure~13). 
This observation further corroborates our trust in the clustering algorithm as a tool to characterise Au nanoparticles melting.
\begin{figure}[b!]
    \centering
    \includegraphics[width=8cm]{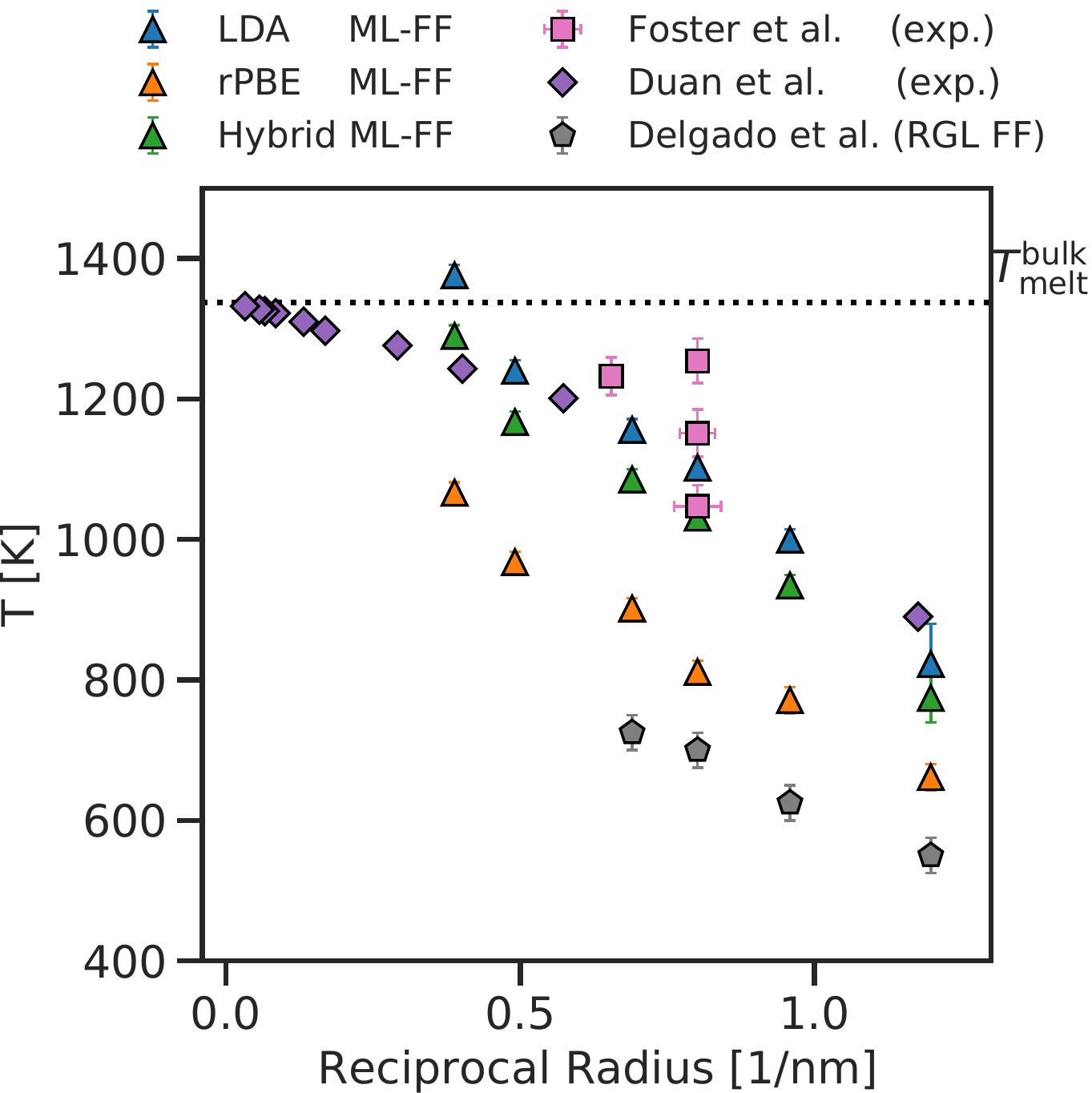}
    \caption{
    \textbf{Melting temperatures of Au NPs of different sizes.}
    Average $T\subt{melt}\supt{NP}$ as a function of NP's reciprocal radius computed for MD simulations employing the LDA-trained (blue triangles), rPBE-trained (orange triangles), and hybrid (green triangles) ML-FFs.
    Experimental data for size-selected Au NPs supported on carbon (pink squares) and spherical Au NPs (purple diamonds), is taken from \citet{Foster2019} and \citet{duan2018influence}, respectively.
    Grey pentagons refer to the $T\subt{melt}\supt{NP}$ estimates from TB-SMA iterative MD melting simulations from \citet{Delgado-Callico2020}.
    Error bars indicate the standard deviation of the melting temperature estimations, and of the NP sizes for experimental data taken from \citet{Foster2019} (pink squares).
}
    \label{fig:melting_temperature}
\end{figure}
\subsection*{Size-dependent melting}
\label{subsec:size_melting}
Figure~\ref{fig:melting_temperature} reports the $T\subt{melt}\supt{NP}$ for NPs of different sizes as a function of the NPs' reciprocal radius, as found during MD simulations carried out with the LDA, rPBE and hybrid ML-FFs. 
The reported $T\subt{melt}\supt{NP}$ is averaged over the 4 (2 for NPs with more than 2500 atoms) independent MD simulations carried out for each NP and each ML-FF.
For an immediate comparison, we report the experimental melting temperature of bulk FCC Au at atmospheric pressure ($T\subt{melt}^{bulk}$ ), and the experimental melting temperatures of Au NPs as a function of the NP size \cite{duan2018influence, Foster2019}.
For reference, we add the $T\subt{melt}\supt{NP}$ estimates obtained using a classical MD where the interatomic interaction is derived in the second-moment approximation of the tight-binding (TB-SMA) \cite{Delgado-Callico2020}.
All the ML-FFs lead to $T\subt{melt}\supt{NP}$ predictions which are (as expected) lower than the ones found during experiments for C-supported Au NPs (pink squares in Figure~\ref{fig:melting_temperature}).
On average, the rPBE-derived ML-FF predicts $T\subt{melt}\supt{NP}$ 250$\pm$50~K lower than the ones predicted by the LDA-derived ML-FF, and 180$\pm$40~K lower than the ones predicted by the hybrid ML-FF.
Interestingly, the $T\subt{melt}\supt{NP}$ predicted by the hybrid ML-FF are less than 50~K away from the melting temperatures found experimentally via differential scanning calorimetry measurements. \cite{duan2018influence}

\subsection*{Melting mechanism characterisation}
\label{subsec:melting_mechanism}
In the previous section we established the quantitative agreement between the ML-FFs' predictions and the experimental melting temperatures of Au NPs, also as a function of their size.
It is then natural to proceed further and analyse the mechanism by which phase changes occur.

\begin{figure}[t!]
    \centering
    \includegraphics[width=9cm]{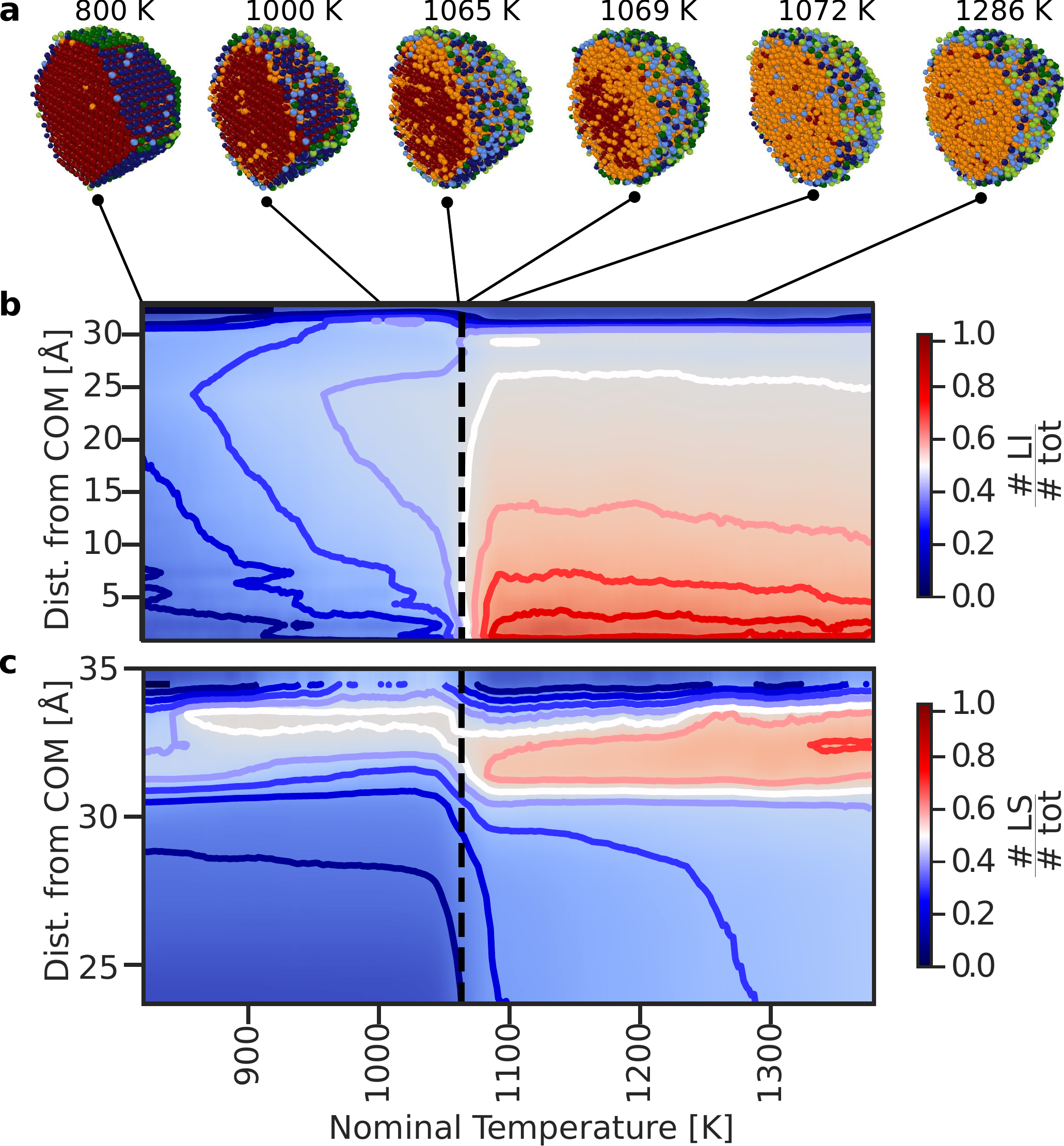}
    \caption{
    \textbf{Distribution of liquid environments in an Au 6266 NP.}}
    Panel a: snapshots of Au 6266 simulated using the r-PBE ML-FF at different nominal simulation temperatures, with atoms coloured according to the clustering algorithm, and using the same colour scheme as in Figure~\ref{fig:maps}.
    Panels b and c: average fraction of \#LI (b) and \#LS (c) local atomic environments as a function of the radial distance from the COM (y coordinate), and of the nominal system temperature (x coordinate).
    The bold coloured lines in panels b and c indicate the isosurfaces in the plot, from 0 to 1 every 0.1, while the black dashed line indicates the $T\subt{melt}\supt{NP}$ of 1065~K.
    \label{fig:cluster_heatmap}
\end{figure}

To this end, we display in Figure~\ref{fig:cluster_heatmap}  example snapshots of an Au 6266 NP at different temperatures (panel a), and the temperature-dependent radial distribution of the fraction of LI (\#LI/\#tot, panel b) and of LS (\#LS/\#tot, panel c) local atomic environments.
The \# symbol indicates the number of atoms belonging to a certain class, where we define:
$\#LS=\#LHS+\#LLS$, and $\#tot=\#LHS+\#LLS+\#LI+\#SHS+\#SLS+\#SI.$
The results we report are found by averaging over the set of independent MD melting simulations employing the rPBE-based ML-FF.
We refer the interested Reader to Supplementary Figures 14 and 15 for the same plots for all systems with 147, 309, 561, 923, 2869 and 6266 atoms and using the three ML-FFs.

In Figure~\ref{fig:cluster_heatmap} {and Supplementary Figures 14 and 15}, the large majority of the local atomic environments are correctly labelled as solid (liquid) at the start (end) of each MD simulation.
The average occurrence of all LI atoms increases with temperature, reaching around 0.5 at the $T\subt{melt}\supt{NP}$ independently of their distance from the NPs' centre of mass (COM).
Areas located few $\text{\AA}$ below the NPs' surface instead display significant abundances of LI atoms also at temperatures below $T\subt{melt}\supt{NP}$.
Such observation is in line with experimental results by \citet{Foster2019}, where a surface melting ($T\subt{melt}\supt{surf.}$) temperature below the $T\subt{melt}\supt{NP}$ was observed for Au NPs of sizes comparable to the ones we analyse.
This $T\subt{melt}\supt{surf.}$ was determined in \citet{Foster2019} by taking the average between the onset temperature for shape changes visible via aberration-corrected scanning transmission electron microscope and the highest temperature for which these did not occur.

%

\begin{figure}[t!]
    \centering
    \includegraphics[width=8cm]{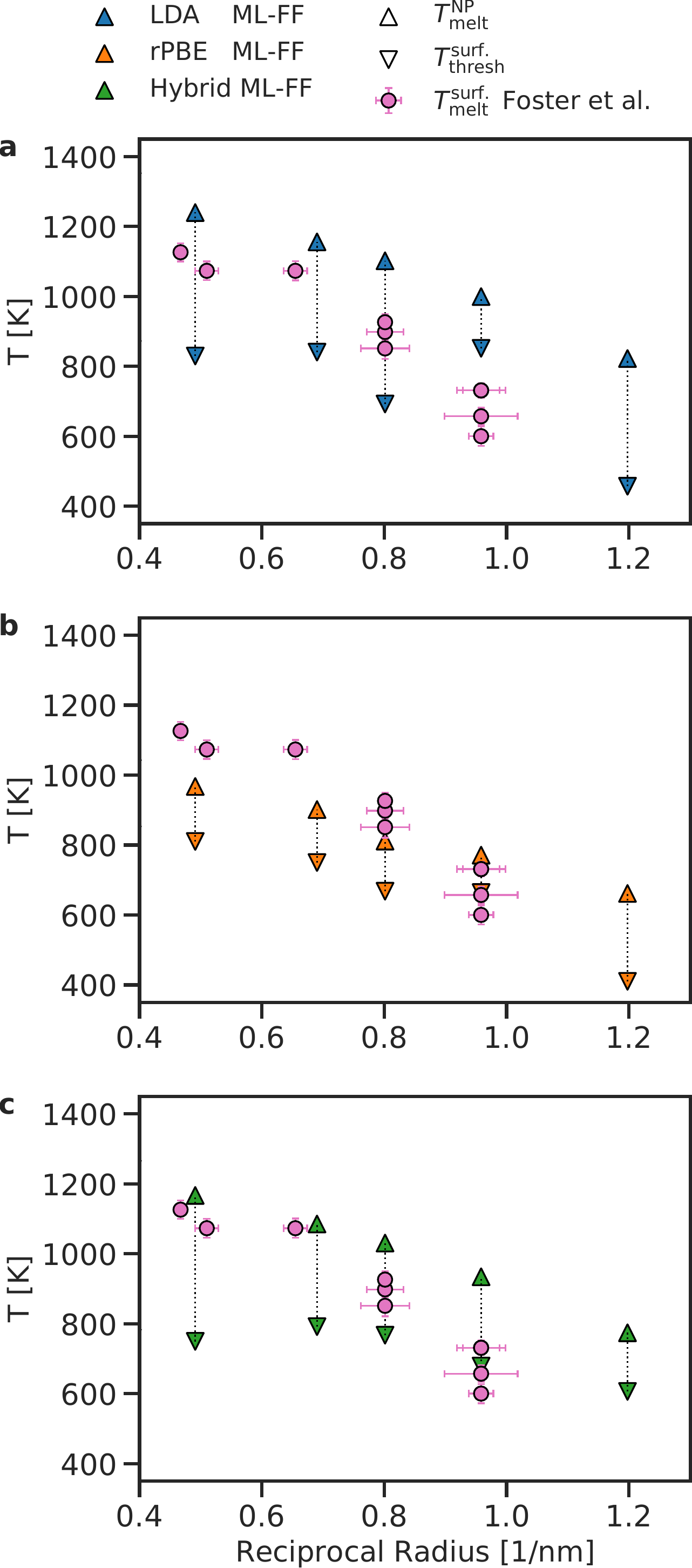}
    \caption{
    \textbf{Surface phase change temperatures of Au NPs of different sizes.}
    $T\supt{surf.}\subt{melt}$  (downward triangles) and $T\subt{melt}\supt{NP}$ (upward triangles) as a function of NPs' reciprocal radius, for MD simulations carried out using the LDA ML-FF (panel a), rPBE ML-FF (panel b), and hybrid ML-FF (panel c).
    Experimental estimates of $T\supt{surf.}\subt{melt}$  from high-resolution TEM measurements are taken from \citet{Foster2019}, and are shown as pink circles.
    Error bars indicate the standard deviation of the melting temperatures and of the NP sizes for experimental data taken from \citet{Foster2019} (pink circles).}
    \label{fig:tmeltsurf}
\end{figure}

To compare our results with available experimental data, we would like to introduce a numerical definition of $T\supt{surf.}\subt{melt}$ .
In analogy to the $T\subt{melt}\supt{NP}$ definition, $T\supt{surf.}\subt{melt}$ should be defined as the temperature at which a clear discontinuity appears in the temperature-dependent evolution of the abundances of liquid-like atoms at the surface of the nanoparticle.
This is, however, not advisable. 
While the number of LI atoms has a clear and distinct positive jump -- which allows us to define a $T\subt{melt}\supt{NP}$ (Supplementary Figure~12) -- 
the temperature-dependent evolution of the number of LS does not show such a clear first-order transition (Supplementary Figure~17).
The relative amount of LS atoms increases gradually with temperature for all NP sizes and all ML-FFs, and reaches values around 0.5 (white line in panel (c) of Figure~\ref{fig:cluster_heatmap} and Supplementary Figures 14 and 15) for atoms in the surface layer at temperatures approaching $T\subt{melt}\supt{NP}$.

We, thus, abandon the search for an unbiased definition of $T\supt{surf.}\subt{melt}$ and introduce the quantity $T\supt{surf.}\subt{thresh}$, which provides an indication of the temperature at which significant surface rearrangement occurs.
The latter is defined as the lowest temperature where at least 0.4 of the local atomic environments in the surface of the NP are classified as liquid (see the Methods section for additional details).
%
%
Figure~\ref{fig:tmeltsurf} reports the values of $T\supt{surf.}\subt{thresh}$ and the $T\subt{melt}\supt{NP}$ for all our MD simulations, and the experimental $T\subt{melt}^{surf.}$ as reported in \citet{Foster2019}.
The temperature ranges comprised between $T\supt{surf.}\subt{thresh}$ and $T\subt{melt}\supt{NP}$ are in line with the experimentally reported $T\subt{melt}\supt{surf.}$, this is especially true for the case of the hybrid ML-FF.

To deepen our understanding of the melting mechanism, we also calculate the mean first-passage temperature (MFPT) required to observe the transition to a liquid phase of 0.4 of the atoms which initially resided at a nanoparticle edge, on a (100) surface, on a (111) surface, or in a bulk environment.
These families of atoms are discriminated against according to their number of first NNs at the beginning of the MD simulations.
Edge atoms (\#NN=6) are more likely to move into a liquid phase than atoms on a (100) facet (\#NN=8), which in turn are more prone to end into a liquid phase than the atoms on a (111) facet (\#NN=9).
The inner atoms (\#NN=12) present the overall largest MFPT. 
This finding is coherent with the melting initiating from the outer layers of the NP (additional details are available in the Methods, and in the Supplementary Methods).

The trends observed during the melting characterization indicate that local phase changes in the outermost layer of a NP start to occur at temperatures a few hundred of K below the $T\subt{melt}\supt{NP}$.
For Au NPs, the proposed characterization protocol establishes that local solid to liquid changes first initiate at low-coordinated atoms at the vertices and edges, then propagate to atoms on (100) and (111) facets, and finally proceed to the inner region of the nanoparticle.

\section*{Discussion}
We characterize the melting mechanism in gold nanoparticles of size 1-6~$nm$, and predict the melting temperatures in good agreement with experimental data using molecular dynamics. 
These simulations employ machine learning force fields,  under the mapped Gaussian process framework, to surpass the trade-off between accuracy and cost in traditional atomistic modelling methods.
We showcase that accurate, efficient, and size-transferable force fields can be trained using small training datasets.
We additionally generate a hybrid 2+3-body force field by linearly combining two machine learning force fields fitted on data computed using different density functional theory functionals; this force field is parametrised to reproduce the bulk cohesive energy and yields predictions of melting temperatures of Au nanoparticles in striking agreement with available experimental data.

To elucidate the melting mechanism, we subsequently develop a general unsupervised clustering approach to differentiate between inner and surface layers and to characterise the phase change at the atomistic level.
Thanks to the insight offered by the proposed clustering algorithm, we demonstrate that the melting transition initiates at the outer layer, and later spreads to the inner region.
The increase in locally liquid environments in the outer region of the nanoparticle before the melting of its core finds a parallel with what is generally referred to in the literature as surface melting.
%
%
The predicted trend is in very good agreement with our experimental observations, where melting was found to start at the outermost layer, at a temperature few hundred of K lower than the NP melting.
We verify that such a melting mechanism occurs regardless of the force-field used to model interatomic interactions, but we also find that different force fields predict different surface and nanoparticle melting temperatures.
We expect that the data-driven simulation and characterisation methods developed here, and the insight we obtain, will stimulate and benefit other research aimed at addressing the complexity of phase changes (solid-to-liquid and liquid-to-solid alike) at the nanoscale.

\section*{Methods}
\label{sec:method}
\subsection*{Database Construction}
\label{subsec:database}
To construct the  training set, we randomly sample 7 frames from a set of 60 frames extracted at regular time intervals from an \textit{ab initio} MD trajectory where an Au NP containing 309 atoms (approx 2 nm in diameter) with an initial FCC morphology undergoes melting from 300~K to 1200~K.
Atomic forces and energy associated with each configuration are calculated within the density functional theory framework, and employing LDA and GGA-rPBE pseudopotentials to generate the training sets for the LDA an rPBE ML-FFs, respectively.
The training sets we employ therefore contain 2163 local atomic environments and associated forces, and 7 total energy values, one for each structure. 
When assessing learning curves (Supplementary Figure~2) we find, in agreement with previous reports \cite{Glielmo2018, Zeni2018}, that the MAE on force prediction converges for training databases which encompass a few hundreds of local atomic environments, and energy predictions do so when energies of a handful of configurations are utilized.
We furthermore note (Supplementary Figure~3) that the shape of the 2-body part of the ML-FFs resulting from training encompassing few hundreds of local atomic environments remains, in essence, unchanged when the number of training points is increased.

We generate a validation set by extracting de-correlated frames from MD trajectories previously reported in \citet{Delgado-Callico2020}, and from \textit{ab initio} MD trajectories previously reported in \citet{Foster2019}.
We sample the melting MD trajectories reported in \citet{Delgado-Callico2020}, carried out using a second-moment tight-binding potential, from 400~K up to 1200~K, and increasing iteratively the temperature of 25~K every 5~ns.
For this setup, we consider NPs containing 146, 147, 192, and 201 atoms which present initial different closed-shell geometries, namely octahedron (146 atoms), icosahedron (147 atoms), Marks decahedron (192 atoms), and regular-truncated octahedron (201 atoms).
The NPs undergo both solid-solid and solid-liquid rearrangements during these MD trajectories (for more details see also the original reference \cite{Delgado-Callico2020}).
The melting MD trajectories reported in \citet{Foster2019} are carried out via NVT simulations, as in the VASP suit, performed at temperatures from 300 to 1200~K with a 150~K interval using LDA DFT, for Au NPs containing 147, 309, and 561 atoms starting from a cuboctahedron.

\subsection*{Machine Learning Force Fields Construction}
\label{subsec:mlff}
We construct the ML-FFs for Au by applying the framework of mappable few-body FFs trained via Gaussian Process regression (GPR) \cite{Glielmo2018, Zeni2018} using the FLARE Python Package. \cite{vandermause2020fly, xie2021bayesian}
GPR FFs hinge on the nearsightedness principle of quantum mechanics to predict total energies for a system of atoms $S$ as a sum of local atomic energy contributions $\varepsilon_i(\rho_i)$:
\begin{equation}
E(S) = \sum_{i \in S} \varepsilon_i(\rho_i),
\label{eq:local_energy}
\end{equation}
where the local atomic energy is predicted as:
\begin{equation}
\varepsilon_i(\rho_i) = \sum_{n} k(\rho_i, \rho_n) \alpha_n.
\label{eq:gpr}
\end{equation}
In Eq.~\ref{eq:gpr}, $k(\rho_i, \rho_n)$ is the kernel (or similarity) function computed between two local atomic environments, the weights $\alpha$ are analytically calculated during the training process, and $n$ is the index that runs from 0 to the number of training data points employed.
We employ 2- and 3-body kernels for local atomic environments, which compare local atomic environments $\rho_i$ based on their distances of pairs and triplets of atoms, respectively. \cite{Glielmo2017, Glielmo2018, Zeni2018}
A local atomic environment $\rho_i$ is defined as the collection of relative positions $r_{ij} = r_j - r_i$ of all atoms $j$ contained within a sphere of radius $r\subt{cut}$ centered on atom $i$.
While traditional GPR FFs are faster to compute than the electronic structure methods they are trained on, they are still orders of magnitude slower than traditional parametrised FFs.
The GPR FFs are therefore transformed into tabulated FFs, which retain the accuracy of the original GPR FFs while being extremely fast to compute, on par with other classical FFs.
The ability to map the GPR FFs follows from the explicit 2- and 3-body nature of the representations we adopt, and takes place via spline interpolation, following the procedure introduced by \citet{Glielmo2018} and first applied to MD simulations in \citet{Zeni2018}.
The hyper-parameters used to train the ML-FFs are, following the notation employed in \citet{vandermause2020fly},  $\sigma_{s, 2}$ = 0.02, $l_{2}$ =  0.4, $\sigma_{s, 3}$ = 7.0, $l_{2}$ =  8.6, $\sigma_n$ = 0.12, $r\subt{cut, 2}$ = 8.0~$\text{\AA}$, $r\subt{cut, 3}$ = 4.5~$\text{\AA}$.

\subsection*{Hybrid ML-FFs}
\label{subsubsec:hybrid}
We generate a third ML-FF, named hybrid, by linearly combining the 2- and 3-body FFs of the ML-FFs derived from LDA and rPBE.
This is done through a parameter $\beta$ that weights the two ML-FFs so that the energy $\varepsilon\supt{hybrid}$ for a local atomic environment $\rho$, is:
\begin{equation}
\varepsilon\supt{hybrid}(\rho) = \beta \varepsilon\supt{LDA}(\rho) + (1-\beta)\varepsilon\supt{rPBE}(\rho).
\end{equation}
The parameter $\beta$ is tuned to match the experimental cohesive energy of bulk Au (3.81~eV/atom) and is set to 0.61 for our ML-FF.
The resulting hybrid ML-FF is a 2+3-body FF, and it has cohesive energy and equilibrium bulk lattice parameter that are intermediate between the LDA and rPBE ML-FFs ones, as can be seen in Supplementary Figure 1.
We remark that the generation of such hybrid ML-FF is possible because of the strictly 2+3-body nature of the ML-FFs employed, and because of the similar functional forms the LDA and rPBE ML-FFs display.
Furthermore, the hybrid ML-FF can be easily fitted to match the experimental cohesive energy of bulk Au solely because this energy is overestimated (underestimated) by the LDA (rPBE) ML-FF.

\subsection*{DFT calculations setup}
\label{subsec:dft}
We employ training data calculated under the Local Density  (LDA) or Generalized Gradient Approximation (GGA - rPBE pseudopotentials) to the exchange-correlation term.
We carry out LDA\cite{Perdew1981} calculations using the Vienna Ab initio Simulation Package\cite{Kresse1996,Kresse1996a} with projector-augmented wave pseudopotentials.\cite{Blochl1994,Joubert1999} 
%
The energy cut-off of the plane-wave basis set was 240~eV, and the tolerance for self-consistency for the electronic steps was set at 10$^{-6}$~eV.
We calculate GGA rPBE \cite{Hammer1999} reference energies and forces using CP2K 6.1. \cite{Hutter2014}
All elements are described with the DZVP-MOLOPT basis set \cite{VandeVondele2007} with cores represented by the dual-space Goedecker-Teter-Hutter pseudopotentials.\cite{Krack2005}
The plane-waves cut-off is set to 500~Ry with a relative cut-off of 50~Ry.
The self-consistent cycle converges when a change of less than 10$^{-6}$~eV is observed in the estimate of the system's energy.

\subsection*{Molecular Dynamics calculations setup}
\label{subsec:md}
To study via ML-FFs the melting of Au NPs, we perform several independent MD simulations at fixed volume in periodic boundary conditions (box width = 100\AA). 
We employ LAMMPS \cite{plimpton1995fast} as our MD engine, and the FLARE \cite{vandermause2020fly} add-on for calculating the energies and forces predicted by the mapped ML-FF.
The temperature of the system, controlled using a Langevin thermostat with a 100~fs noise, continuously increases at a rate of 20~K/ns, with starting temperatures that range between 400K and 700K and ending temperatures that range between 1200K and 1500K, depending on the NP's sizes.
Newton's equations of motions are integrated via a velocity-Verlet algorithm with a 1~fs time step for systems with less than 1000 atoms, and 2~fs for systems above 1000 atoms in the case of the LDA and rPBE ML-FFs.
All simulations employing the hybrid ML-FF are carried on using a 5~fs integration time step.
\subsection*{Local Atomic Environment Descriptor}
\label{subsubsec:descriptor}
We employ a local atomic density descriptor to feature each atomic environment in a NP as a function of the relative positions of the other atoms within a cut-off set to 1.75 times the average nearest neighbour distance, and therefore set to 4.24~$\text{\AA}$ for simulations employing the LDA ML-FF, to 4.42~$\text{\AA}$ for simulations employing the r-PBE ML-FF, and to 4.30~$\text{\AA}$ for simulations employing the hybrid ML-FF.
A sensitivity analysis shows that the featurisation associated with the representation is marginally affected by the choice of the cut-off radius, as long as the latter is larger than the bulk second nearest neighbours distance (see Supplementary Methods for further detail). 
We adopt the 2+3-body atomic cluster expansion representation with 4 radial and 4 angular components and employ Bessel functions of the first kind as radial basis functions. \cite{Drautz2019,Drautz2020,Zeni2021}
\subsection*{Clustering Algorithm for Phase Change Characterisation}
\label{subsubsec:clustering}
To apply the clustering algorithm to data generated through the use of a ML-FF, we first gather 10000 randomly chosen local atomic environment representations from among MD simulations of all NP sizes.
We then run a hierarchical k-means clustering  \cite{Macqueen1967} algorithm to group similar representations, applying two to three iterations of k-means clustering to partition the local atomic environment sampled during the MD simulations into the six classes described previously (additional details can be found in the Supplementary Methods).

\subsection*{Melting Temperature Estimation}
\label{subsubsec:melting_t}
We estimate the $T\subt{melt}\supt{NP}$ as the temperature for which the maximum positive derivative of the fraction of inner atoms labelled as liquid w.r.t. the nominal simulation temperature (or, equivalently, the simulation time) is observed.
The $T\subt{melt}\supt{NP}$ is commonly defined as the temperature where the highest value of the heat capacity is observed $T\subt{melt}\supt{NP}$, or as the temperature here the highest standard deviation in the total energy is found. \cite{Delgado-Callico2020, chen2020heating, dai2017test}
Supplementary Figure~13 shows the striking correspondence that exists between the $T\subt{melt}\supt{NP}$ estimated using the three aforementioned methods.
This result confirms that the $T\subt{melt}\supt{NP}$ estimation methods we introduce are accurate for the systems we consider and reinforces our belief that the characterization offered by our clustering method is valid.

\subsection*{Surface Transition Temperature Calculation}
\label{subsubsec:surface_t}
\begin{figure}[t!]
    \centering
    \includegraphics[width=7cm]{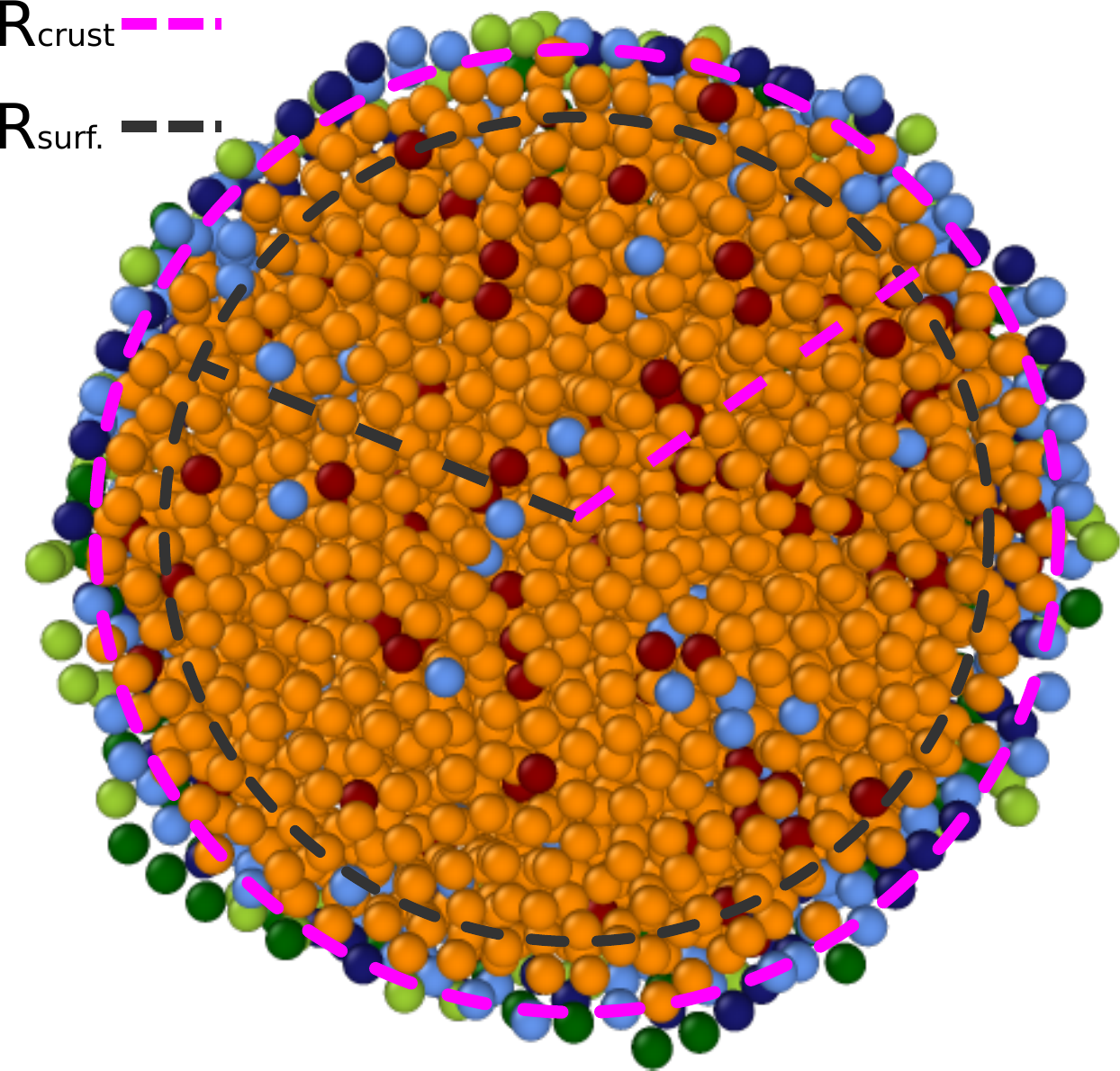}
    \caption{
    \textbf{Crust and surface radiuses in an Au 6266 NP.}}
    Depiction of the values of $R\subt{crust}$ (pink) and $R\subt{surf.}$ (grey) for an Au 6266 NP.
    Atoms are colour-coded according to their class as yielded by the clustering algorithm, and mirror the ones of Figures \ref{fig:maps} and \ref{fig:cluster_heatmap}.
    \label{fig:Au_slice}
\end{figure}

To calculate $T\supt{surf.}\subt{thresh}$, we analyse the spatial distribution of LS atoms.
We subdivide the NP in spherical shells of width 1 \AA ~ centered at the COM of the NP.
We define the crust radius, $R\subt{crust}$, as the distance from the NP COM of the spherical shell where the highest fraction of LS atoms resides.
This generally coincides with the outermost radial shell of atoms in the NP.
We then aim to define a surface shell, and consider a second distance, $R\subt{surf.} = R\subt{crust} - 3~\text{\AA} $.
The choice of a 3~$\text{\AA}$ buffer represents an arbitrary but educated guess to incorporate, approximately, a second shell of atoms in our statistics.
Finally, we define $T\supt{surf.}\subt{thresh}$ as the lowest temperature at which the liquid local atomic environments in the surface of the NP amount for the 0.4 of the total number of local atomic environments in the surface shell.
To exemplify the protocol,  Figure~\ref{fig:Au_slice} displays the values of $R\subt{surf.}$ and $R\subt{crust}$ for a snapshot extracted from an MD trajectory sampled using the rPBE ML-FF.

\subsection*{Mean First-Passage Temperature}
\label{subsubsec:mfpt}
To evaluate the mean first-passage temperature (MFPT), we monitor the label assigned to each atom in the system by the hierarchical clustering scheme at each time step.
The MFPT is then defined as the lowest temperature at which at least 0.4 atoms of given initial coordination are labelled as  liquid environments.
Since MFPTs depend on the $T\subt{melt}\supt{NP}$, for all MD trajectories we normalize each MFPT by the average $T\subt{melt}\supt{NP}$ for that particular NP size and ML-FF employed.

\subsection*{Statistical Information}
\label{subsec:statistical_info}
Simulation results are obtained as averages over 4 independent simulations for NPs containing less than 1000 atoms, and over 2 independent simulations for NPs containing more than 1000 atoms.
The $T\subt{melt}\supt{NP}$ reported for each NP size and ML-FF are the average $T\subt{melt}\supt{NP}$ computed across the 4 (or 2) independent MD simulations.
The error bars for $T\subt{melt}\supt{NP}$ reported for the y-axis of Figure~\ref{fig:melting_temperature} and Supplementary Figures 6 and 13, are calculated as the maximum between 25~K - the temperature window (see also Methods) used to individuate the peak of the positive derivative of the fraction of liquid atoms w.r.t. simulation temperature - and the standard deviation of $T\subt{melt}\supt{NP}$ computed for the 4 (or 2) independent MD simulations for each NP size and ML-FF used to simulate them.
The mean absolute errors (MAEs) on energy differences (force components) reported in Supplementary Table~1 are computed on a variable number of observations, determined by the NP size, from 9 (15147) for Au 561 Co to 50 (22050) for Au 147 Co.
On average, MAEs on energy differences (force components) are calculated on 33~$\pm$~14 (21417~$\pm$~12531) samples.

\section*{Data Availability}
The tabulated Au ML-FFs, Au NPs MD trajectories, and ab initio training data for Au NPs generated in this study have been deposited in the Materials Cloud database under accession code
\href{https://archive.materialscloud.org/record/2021.131}{https://archive.materialscloud.org/record/2021.131}. \cite{materialscloud}
Example MD trajectories are also stored in the same repository.
Source data for Figures~\ref{fig:melting_temperature} and \ref{fig:tmeltsurf} are provided with this paper.
Other data are available from the authors upon request.

\section*{Code availability}
A majority of the code used in this calculation is open source.
ML-FF training and mapping are carried on using FLARE (https://github.com/mir-group/flare). 
DFT data are gathered using CP2K (https://www.cp2k.org) and VASP (https://www.vasp.at - licence number 5-867). 
MD Simulations are run via LAMMPS (https://lammps.sandia.gov).
The computation of local atomic environment descriptors, and the clustering characterization are carried on using the Raffy Python package \cite{Raffy}.
K-means clustering is done in Python via the SciPy library.

\section*{References}

\bibliography{main_arxiv}

\section*{Acknowledgments}
\label{sec:acknowledgments}
C.Z acknowledge funding by the Engineering and Physical Sciences Research Council (EPSRC) through the Centre for Doctoral Training Cross-Disciplinary Approaches to Non-Equilibrium Systems (CANES, Grant No. EP/L015854/1) and by the European Union’s Horizon 2020 research and innovation program (Grant No. 824143, MaX `MAterials design at the eXascale' Centre of Excellence).
K.R.  has received funding from the European Research Council (ERC) under the European Union’s Horizon 2020 research and innovation programme, (Marie Curie Individual Fellowship Grant agreement No. 890414).
S.dG. acknowledges funding from European Union’s Horizon 2020 research and innovation program (Grant No. 824143, MaX MAterials design at the eXascale Centre of Excellence). 
F.B. acknowledges the financial support offered by the Royal Society under project number RG120207 and DIPC for supporting her visiting professorship.
We are grateful to the UK Materials and Molecular Modelling Hub for computational resources, partially funded by EPSRC (EP/P020194/1 and EP/T022213/1), our membership of the Materials Chemistry Consortium, funded by EPSRC (EP/R029431), the Swiss National Supercomputer Center (CSCS) 
(project “sm54”), and to the Supercomputing Wales project, partially funded by the European Regional Development Fund (ERDF) via the Welsh Government. 
\section*{Author contributions}
\label{sec:author_contributions}
F.B, T.P. and R.E.P. originally envisioned the project. 
C.Z., K.R., and F.B. designed and interpreted the simulations and their post-processing.
C.Z. carried on the machine learning force-field fitting and performed the ML-FFs melting simulations. 
C.Z. implemented the characterisation algorithm.
K.R. and T.P. performed the electronic structure calculations.
C.Z., K.R., T.P, J.K., S.d.G, R.E.P. and F.B. contributed to the writing of the paper.
\section*{Competing Interests}
\label{sec:competing_interests}
The authors declare no competing interests.
\section*{Supplementary Methods}
\label{subsec:supplementary_materials}
The Supplementary Methods contain further details on the ML-FF training, validation and volume-energy curves, on the hierarchical clustering procedure we develop, and on the melting and surface phase changes we observe during the MD simulations.
\newpage
\clearpage

\renewcommand{\thetable}{S\arabic{table}}
\renewcommand{\thefigure}{S\arabic{figure}}

\setcounter{figure}{0}

\section*{Supplementary Figures}
\begin{figure}[h!]
    \centering
    \includegraphics[width = 8cm]{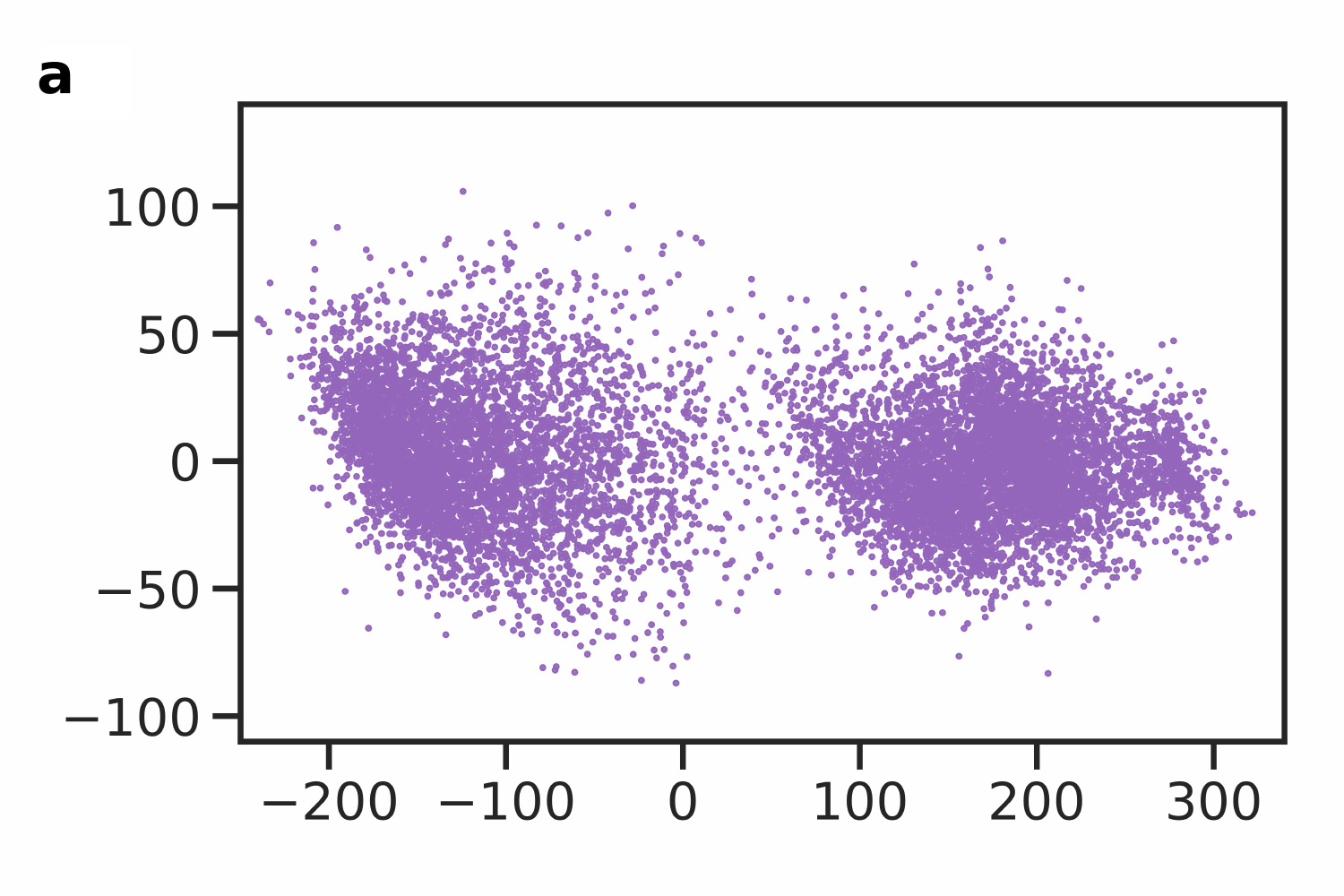}
    \includegraphics[width = 8cm]{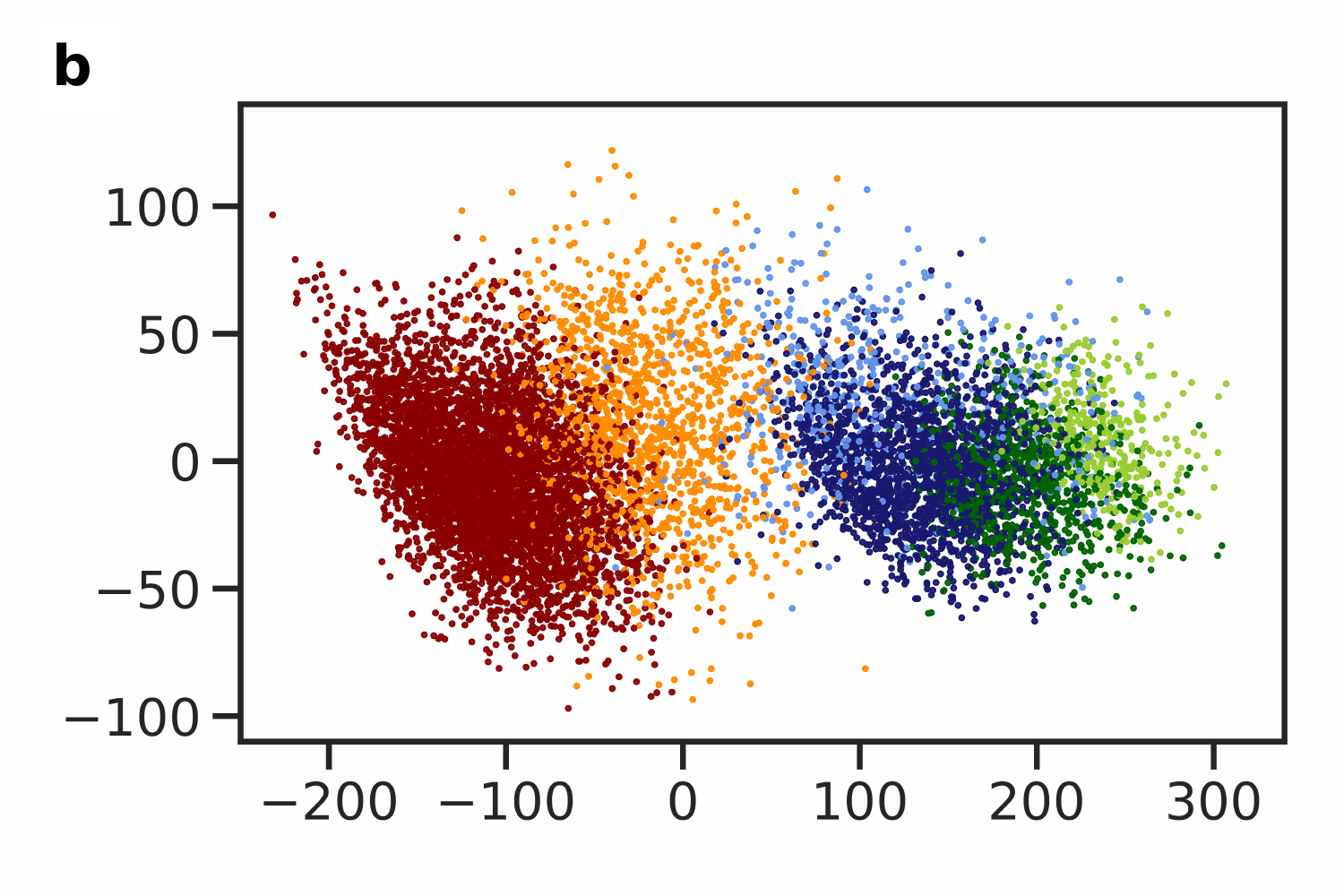}
    \caption{
    \textbf{Visualization of the heterogeneity of the training dataset.}
    2-dimensional PCA projection of the 40-dimensional atomic cluster expansion descriptor computed on (panel a) 10000 local atomic environments sampled from the Au 309 MD simulation used to generate the initial training set for the LDA ML-FF, and (panel b) from 10000 local atomic environments sampled from MD simulations of Au147, 309, 561, 923, 2869, and 6266 carried out using the LDA ML-FF.
    Points in panel b are also colour-coded according to their classification using the hierarchical k-means clustering algorithm, and following the same colour scheme as in Figure 1 and Supplementary Figures~\ref{fig:maps_lda} and \ref{fig:maps_hybrid}}.
    \label{fig:db}
\end{figure}
\begin{figure}[!htb]
    \centering
    \includegraphics[width = 8cm]{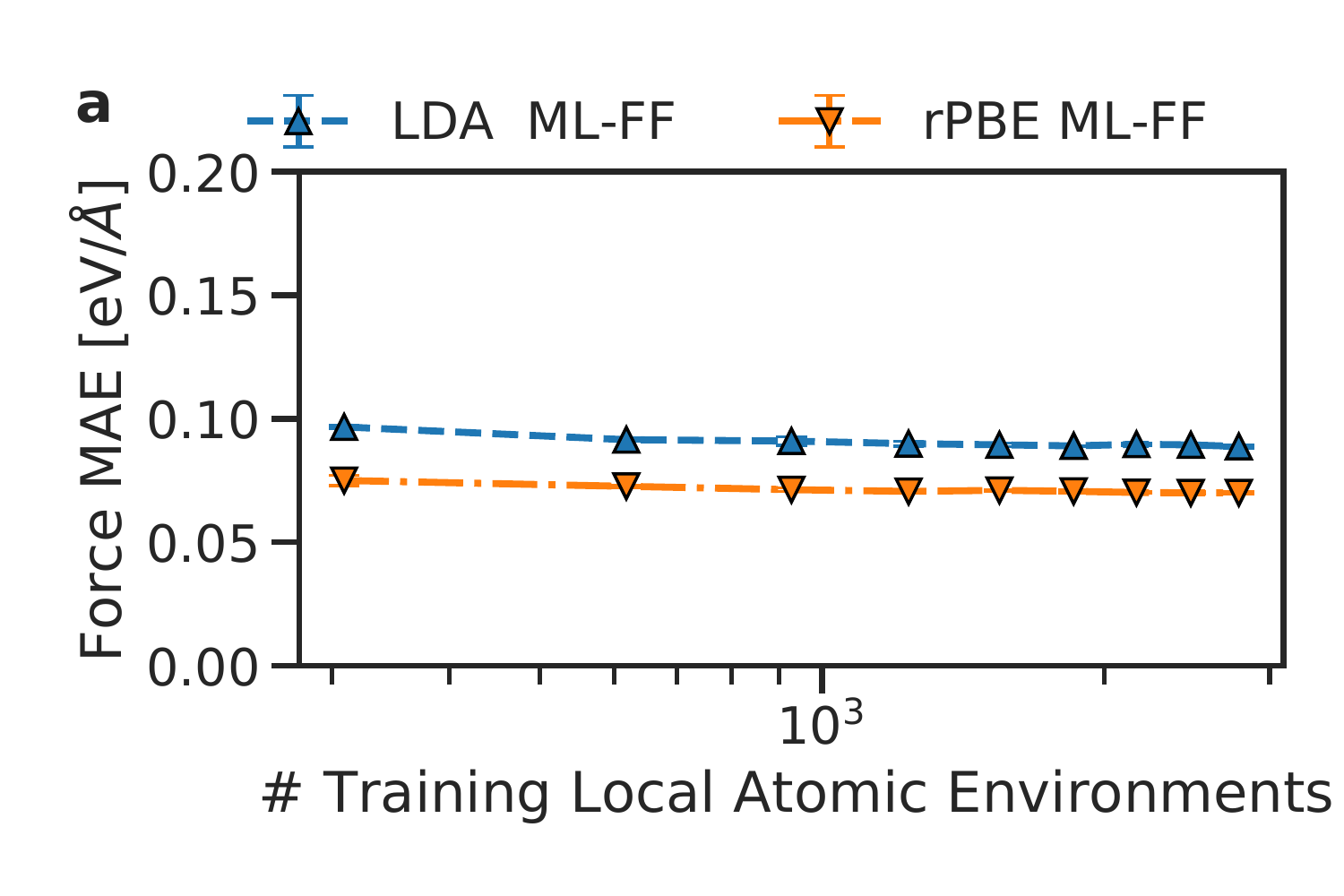}
    \includegraphics[width = 8cm]{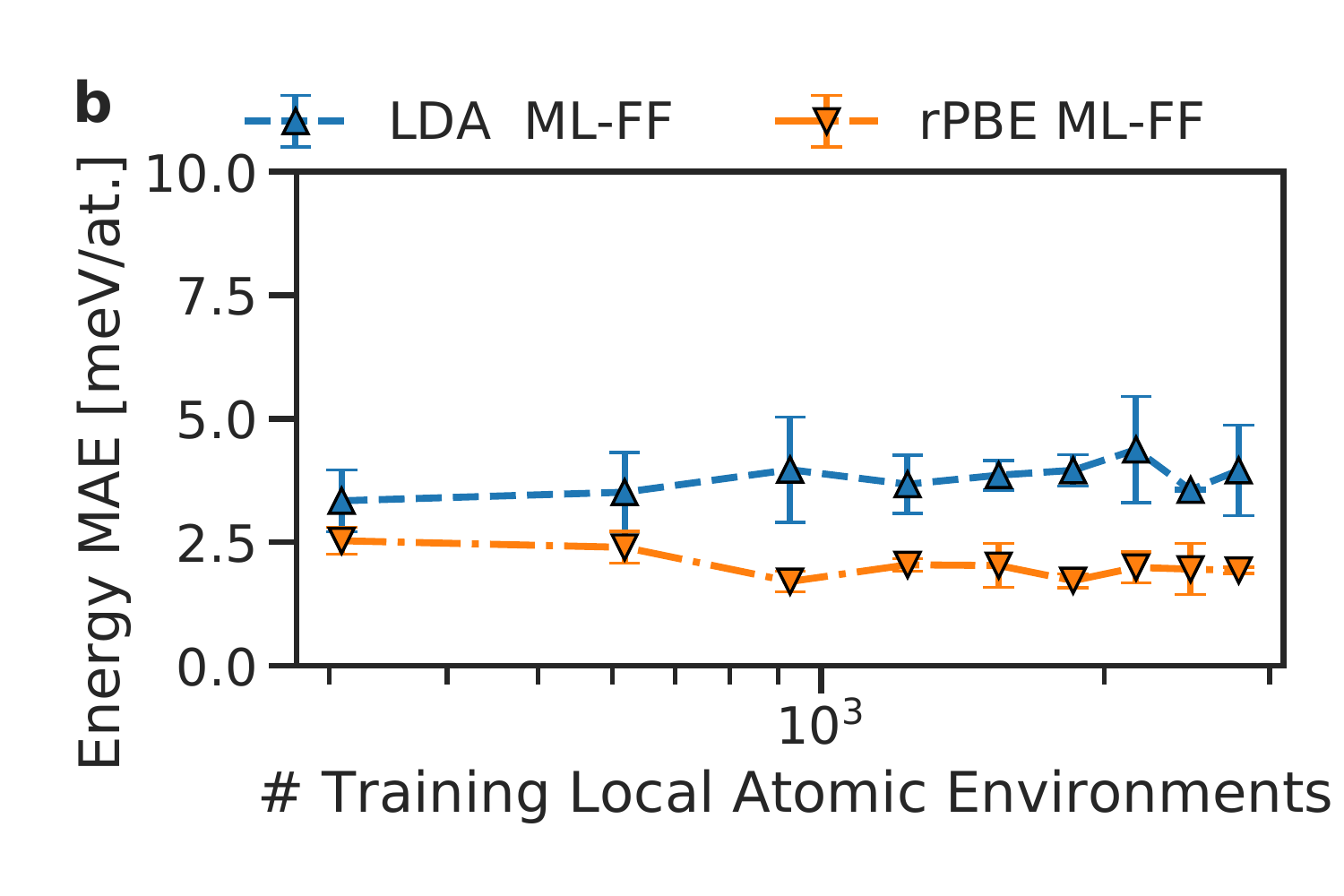}
    \caption{
    \textbf{Training curves for the LDA and rPBE ML-FFs.}
    Test MAEs incurred by the LDA (blue) and rPBE (orange) ML-FFs on force components (panel a) and atomic energy differences (panel b), as a function of the number of local atomic environments used to train them.
    The error bars represent the variance of the MAEs across three independent training and testing iterations.}
    \label{fig:learning_curves}
\end{figure}
\begin{figure}[!htb]
    \centering
    \includegraphics[width = 7.5cm]{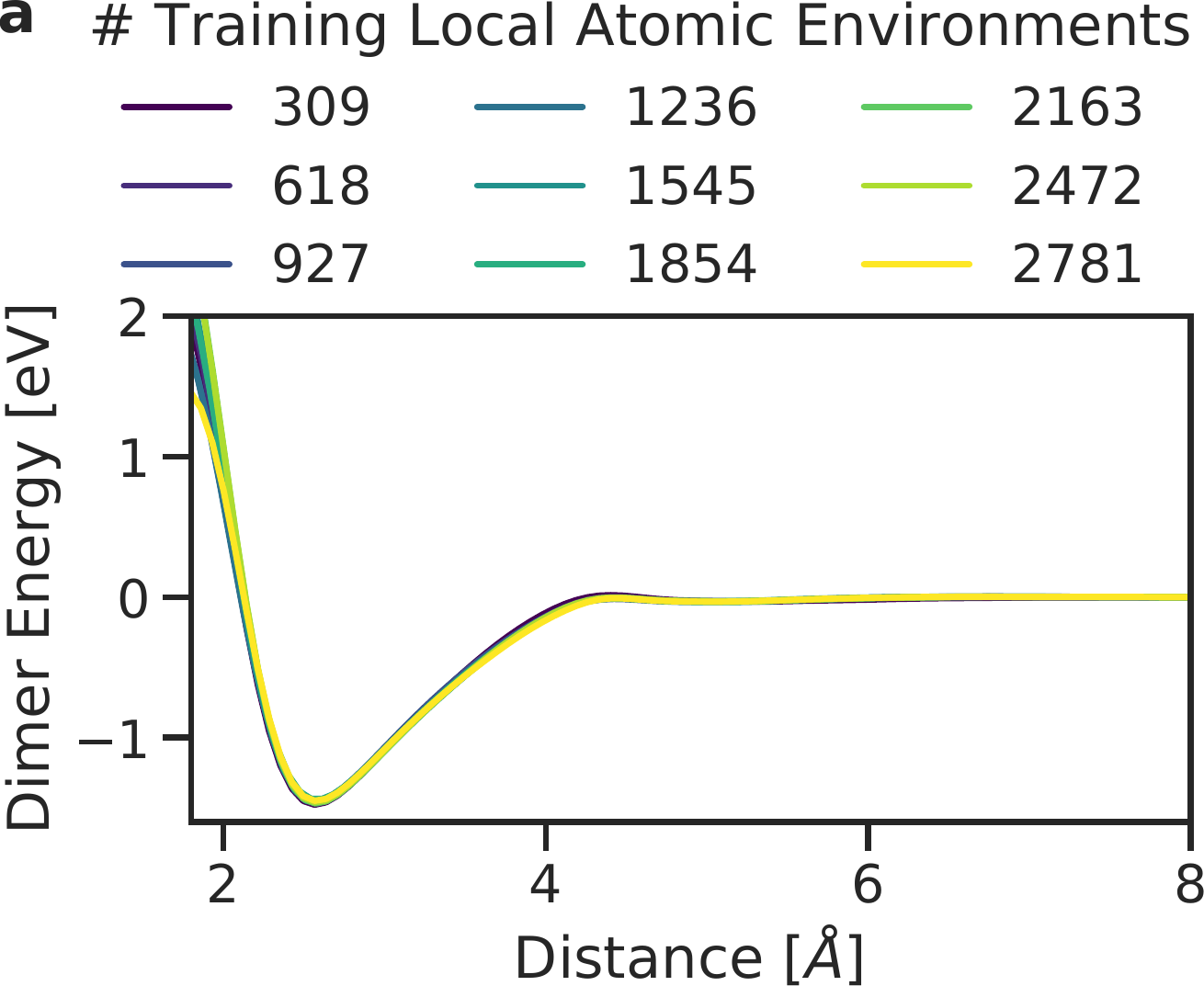}
    \hspace{1cm}
    \includegraphics[width = 7.5cm]{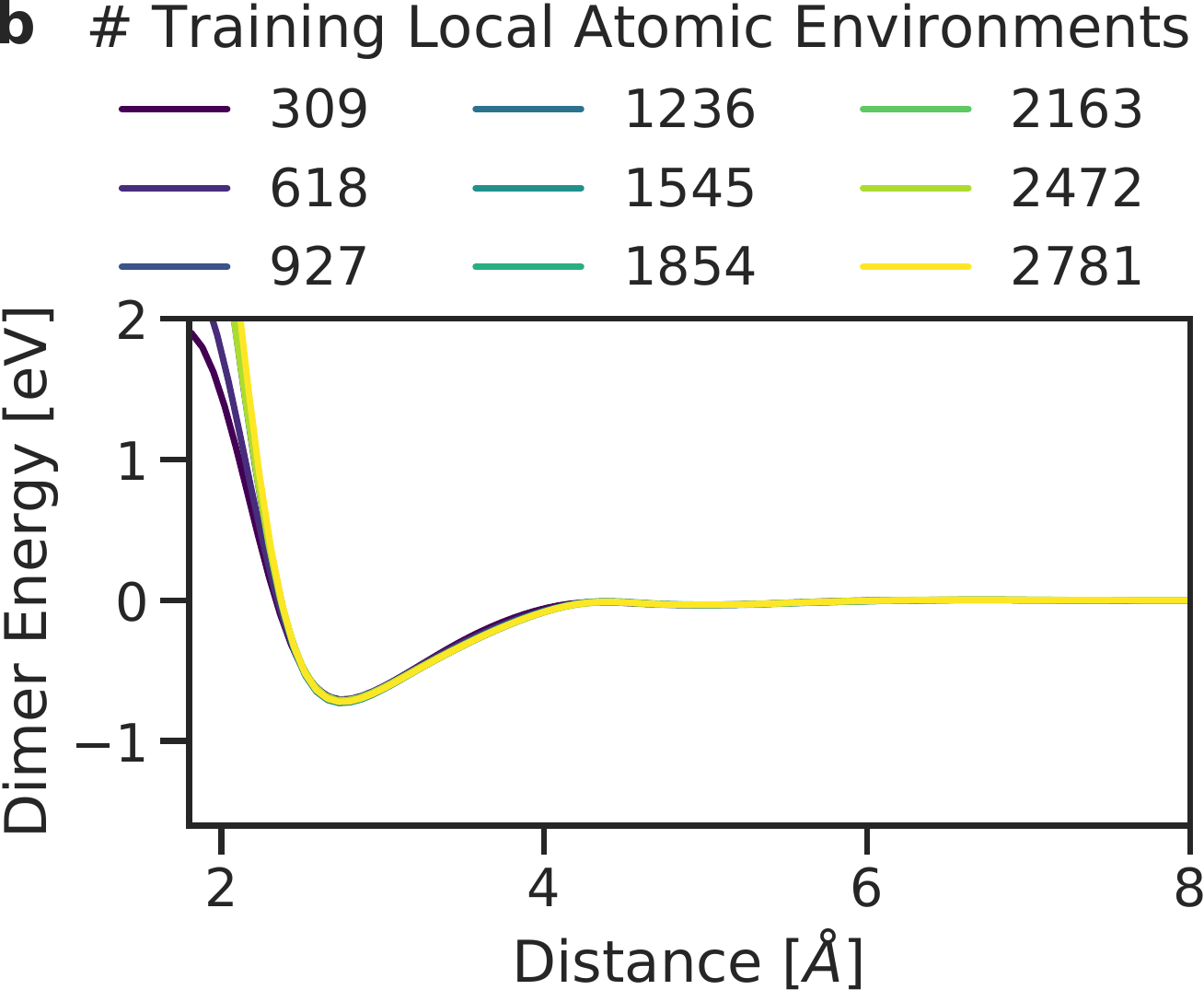}
    \caption{
    \textbf{Dimer energy for LDA and rPBE ML-FFs using different training set sizes.}
    Dimer energy as a function of distance for LDA (panel a) and rPBE (panel b) ML-FFs trained on an increasing number of local atomic environments (color-coded, from blue to yellow).}
    \label{fig:dimer}
\end{figure}
\begin{figure}[!htb]
    \centering
    \includegraphics[width = 8cm]{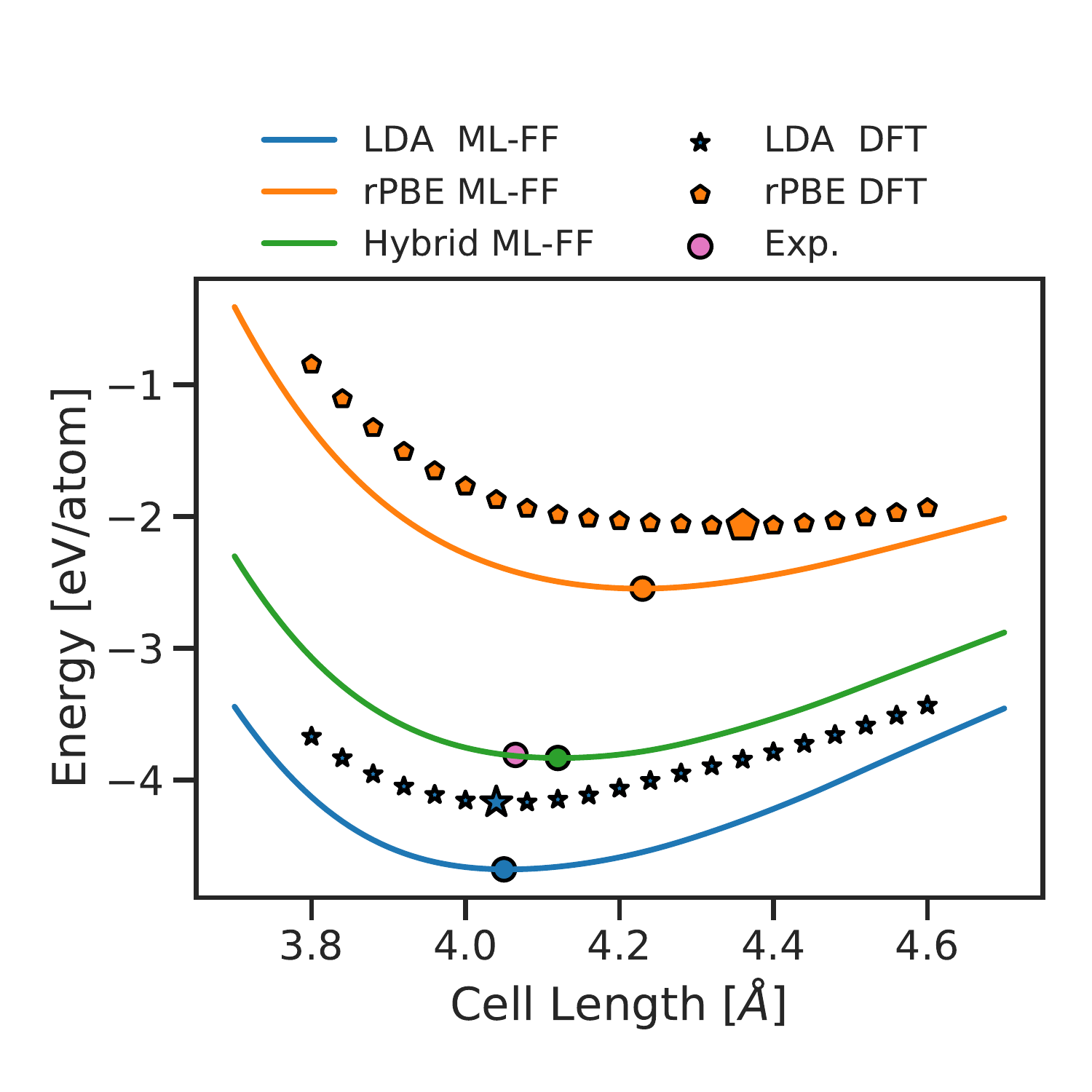}
    \caption{
        \textbf{Volume-energy curves for the three ML-FFs and for LDA and rPBE ab initio calculations.}
        Cohesive energy per atom as a function of cell length in FCC Au computed using the LDA (blue), rPBE (orange), and hybrid (green) ML-FFs, and via ab initio calculations carried out using the LDA (blue stars) and rPBE (orange pentagons) psudopotentials used to generate the respective training sets.
    Blue, orange and green dots indicate the equilibrium cell length at zero pressure for the three ML-FFs, and a large orange pentagon and a large blue star indicate the same values for the DFT calculations, while a pink dot indicates the experimental cohesive energy \cite{kittel2005crystal} per atom and cell length \cite{davey1925precision} of FCC Au, for reference.}
    \label{fig:volume_energy_curves}
\end{figure}
\begin{figure*}[!htb]
    \centering
    \includegraphics[width=17.5cm]{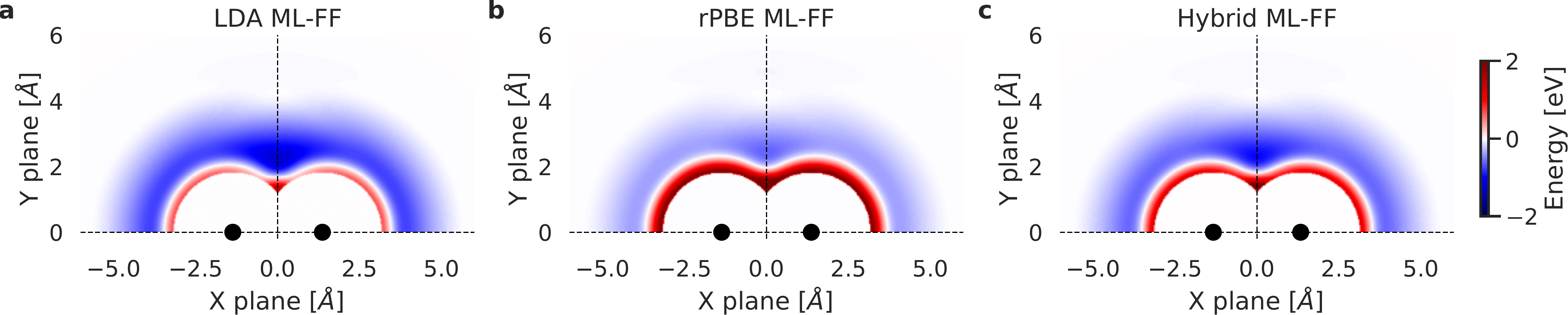}
    \caption{
    \textbf{Potential energies of three Au atoms on the plane for the three ML-FFs.}
    Potential energy (in colour) of an Au atom neighbouring two Au atoms (black dots) as a function of its position in the x-y plane for two M-FFs trained on ab initio data extracted from LDA (a) and r-PBE (b) DFT simulations, and for the hybrid Ml-FF (c). 
    The two neighbouring atoms are lying on the x-axis and their distance is the distance of minimum energy for an Au dimer for that FF.
    The potential energy felt by the Au atom is a sum of 2- and 3-body contributions, and is displayed for interatomic distances $> 1.8$ $\text{\AA}$.}
    \label{fig:pes}
\end{figure*}
\begin{figure}[!htb]
    \centering
    \includegraphics[width = 8cm]{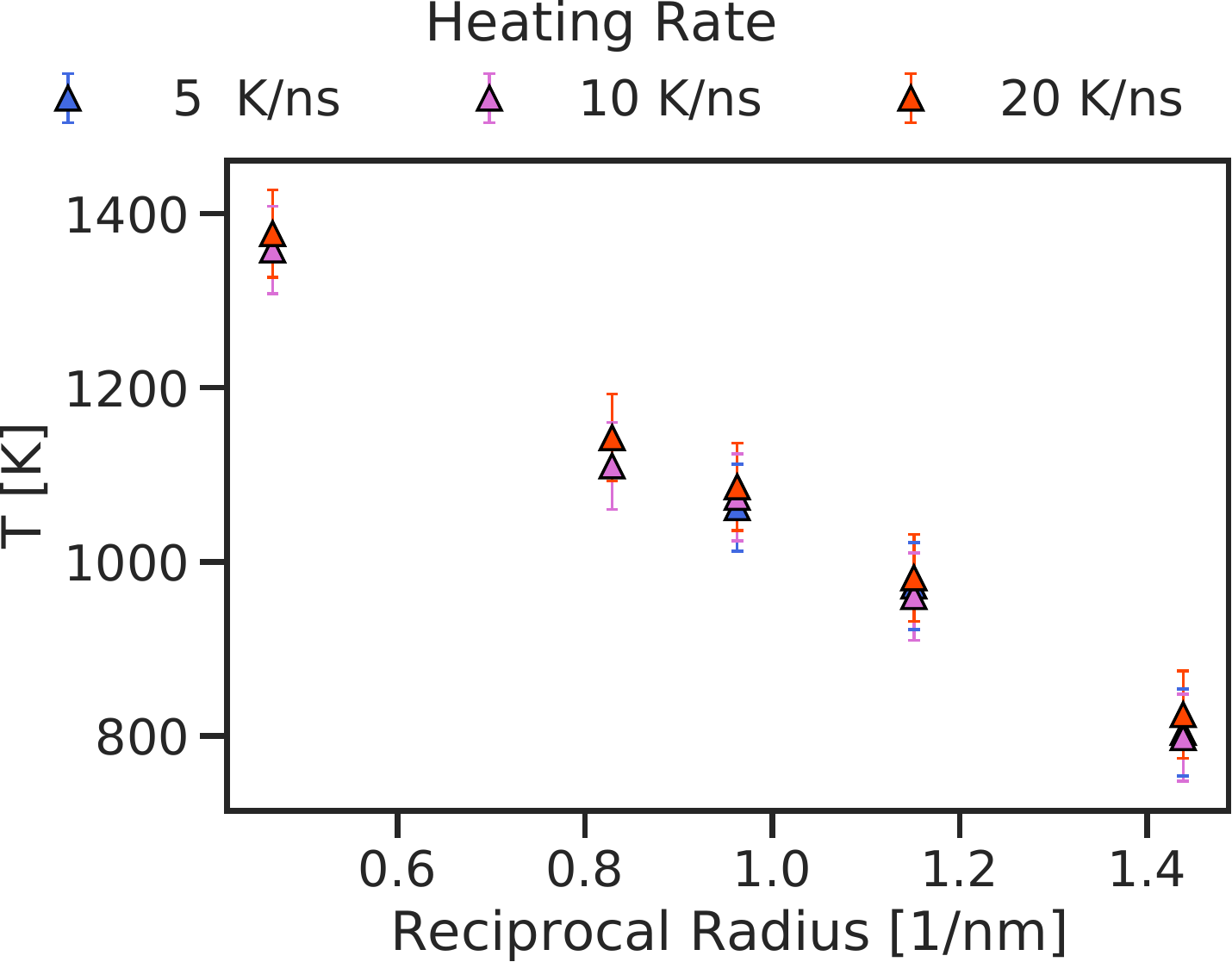}
    \caption{
    \textbf{Effect of heating rate on melting temperature of Au NPs.}
    $T\subt{melt}\supt{NP}$ as a function of inverse NP radius for MD simulations of Au 147, 309, 561, 923, and 6266 carried out using the LDA ML-FF and with different heating rates (in colour).
    The error bars report the maximum between the standard deviation of the melting temperatures across 4 independent simulations (2 for Au 6266), and 25~K, the temperature averaging window used to estimate the melting temperature.}
    \label{fig:heating_rate}
\end{figure}
\begin{figure*}[!htb]
    \centering
    \includegraphics[width=17cm]{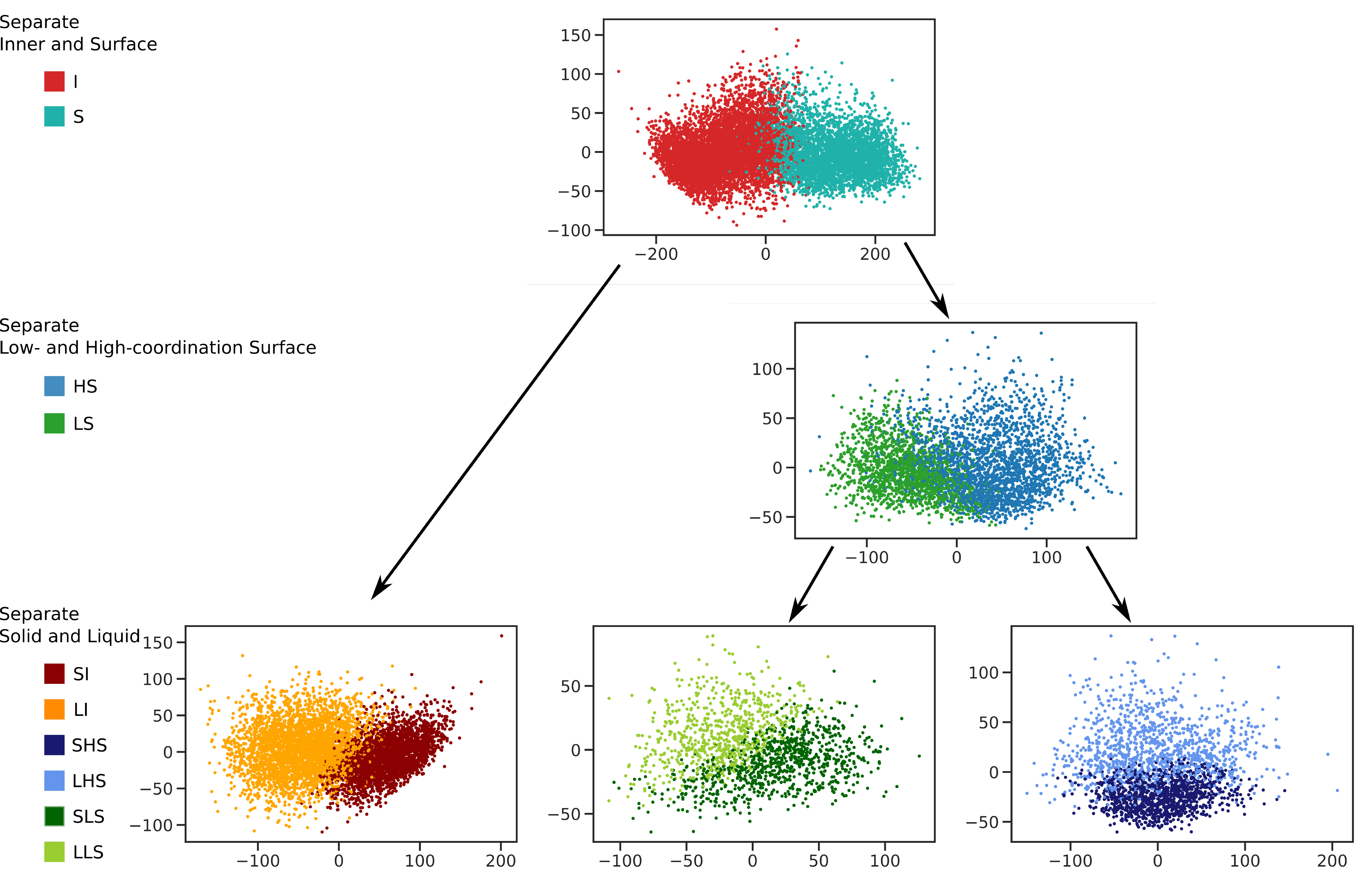}
    \caption{\textbf{Schematic depiction of the hierarchical k-means clustering algorithm on local atomic environments.}
    The plots report different 2-dimensional PCA projections of 10000 local atomic environment descriptors sampled randomly from MD melting simulations of Au NPs containing 14, 309, 561, 923, 2869, and 6266 atoms carried out using the rPBE ML-FF.
    From top to bottom, the clustering classifies the local atomic environments as inner (bright red) and surface (green-blue).
    The second clustering, applied only to surface local atomic environments, is then used to distinguish between high- (blue) and low- (green) coordination local atomic environments.
    The third and last clustering is used to separate the liquid (light colours) from the solid (dark colours) local atomic environments, both in inner, high-coordination, and low-coordination local atomic environments.
    }
    \label{fig:pca1}
\end{figure*}
\begin{figure*}[!htb]
    \centering
    \includegraphics[width=17cm]{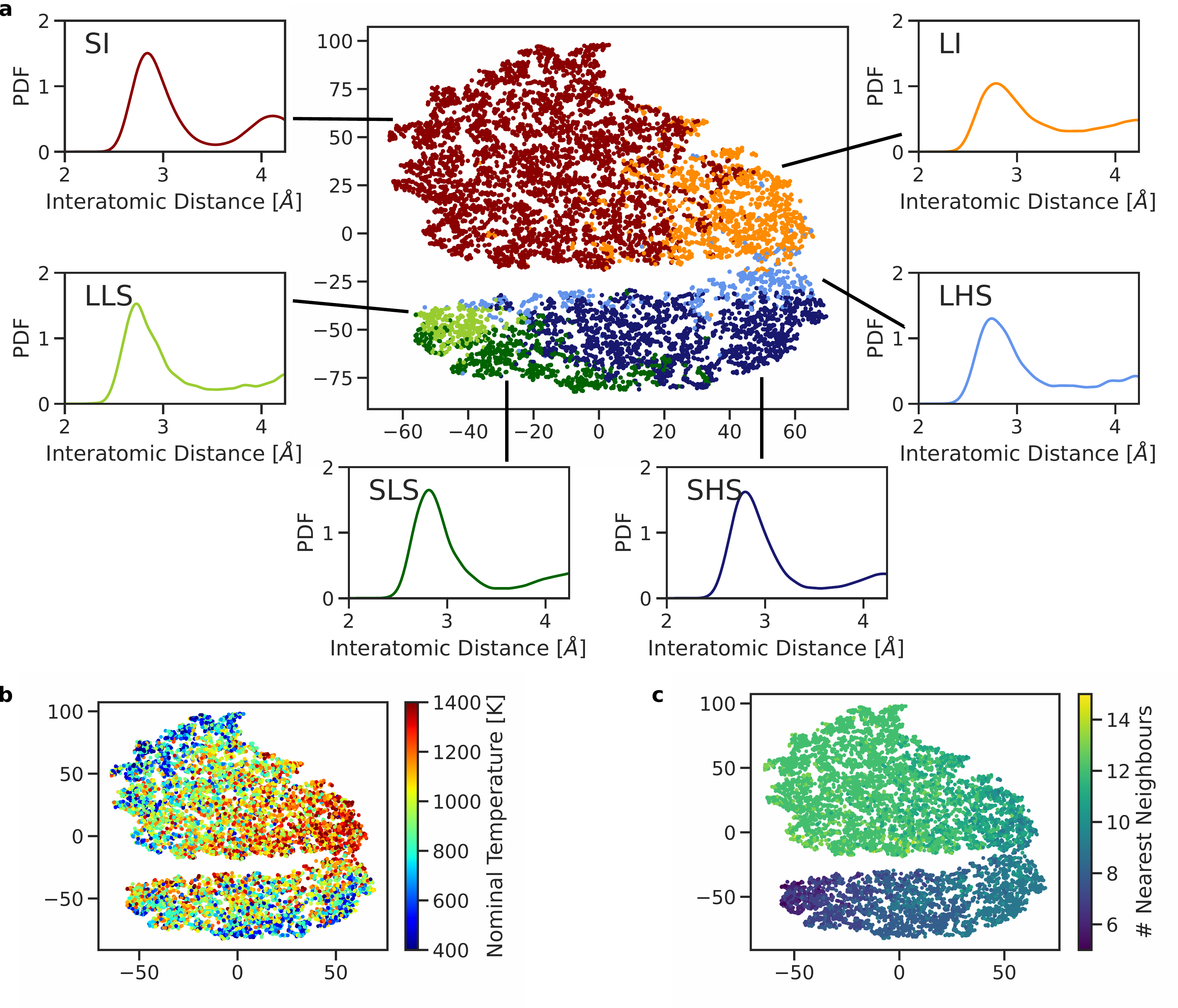}
    \caption{
    \textbf{Six classes of local atoms environments identified through clustering of MD simulations using the LDA ML-FF.}
    Visualization of the local atomic environment representation hierarchical k-means clustering results for MD simulations of Au nanoparticles with 147, 309, 561, 923, 2869 and 6266 atoms, carried out using the ML-FF trained on LDA-DFT data.
    a) 1st and 2nd component (x- and y-axis) of the t-sne projection of the atomic expansion coefficients of 10$^4$ local atomic environments randomly sampled from melting MD simulations. 
    The colours label the six classes assigned by the hierarchical k-means clustering algorithm, as defined in the main text.
    The normalized average pair-distance distribution function (PDF) belonging to each class is also reported. 
    b), c) Same t-sne projection as in a), with colours indicating the nominal simulation temperature at which the local environment was taken from in b), and the number of nearest neighbours using a $r\subt{cut}$ of 3.4 $\text{\AA}$ in c).
    }
    \label{fig:maps_lda}
\end{figure*}
\begin{figure*}[!htb]
    \centering
    \includegraphics[width=16cm]{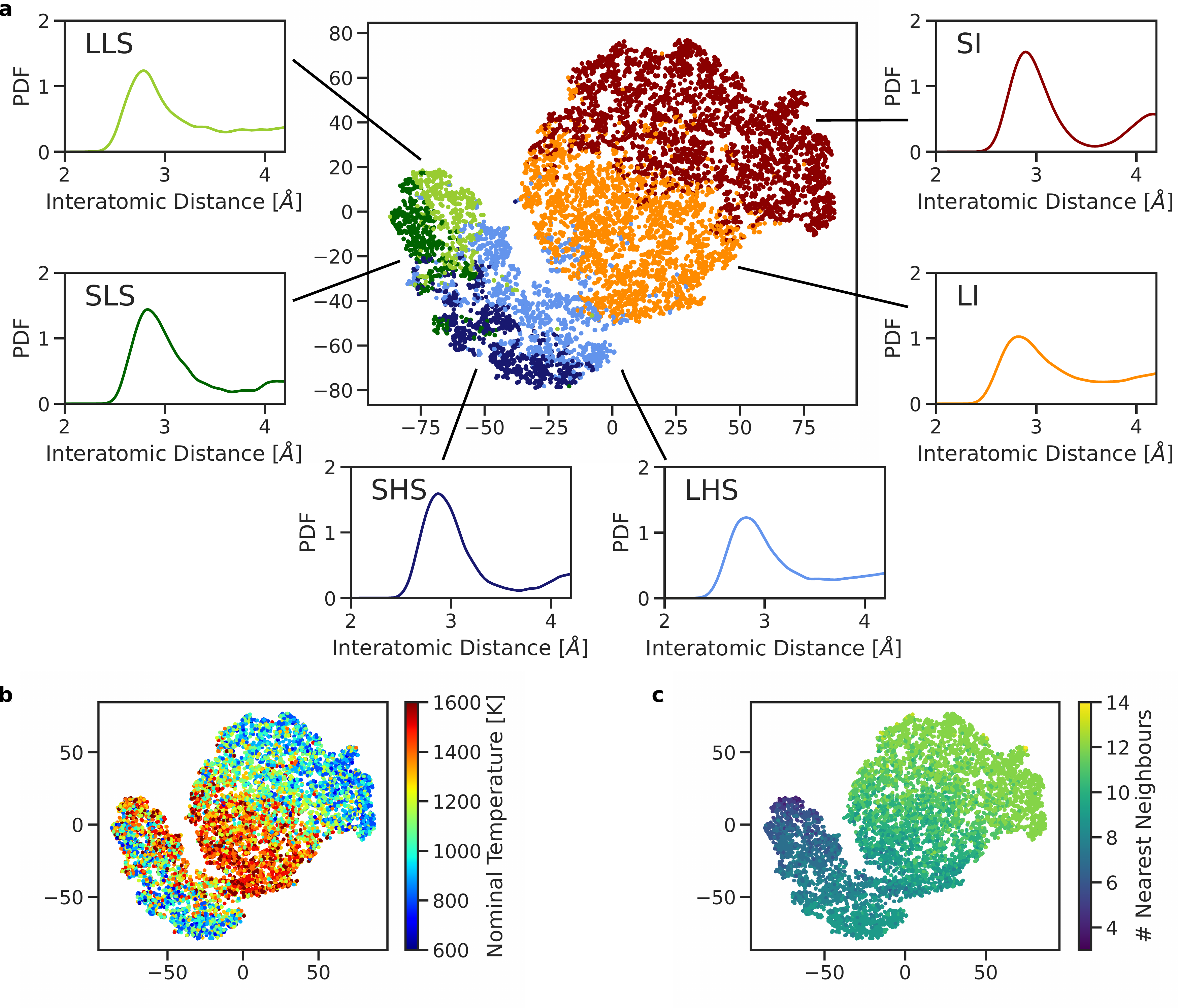}
    \caption{
    \textbf{Six classes of local atoms environments identified through clustering of MD simulations using the hybrid ML-FF.}
    Visualization of the local atomic environment representation hierarchical k-means clustering results for MD simulations of Au nanoparticles with 147, 309, 561, 923, 2869 and 6266 atoms, carried out using the hybrid ML-FF
    a) 1st and 2nd component (x- and y-axis) of the t-sne projection of the atomic expansion coefficients of 10$^4$ local atomic environments randomly sampled from melting MD simulations. 
    The colours label the six classes assigned by the hierarchical k-means clustering algorithm, as defined in the main text.
    The normalized average pair-distance distribution function (PDF) belonging to each class is also reported. 
    b), c) Same t-sne projection as in a), with colours indicating the nominal simulation temperature at which the local environment was taken from in b), and the number of nearest neighbours using a $r\subt{cut}$ of 3.4 $\text{\AA}$ in c).
    }
    \label{fig:maps_hybrid}
\end{figure*}
\begin{figure*}[b!]
    \centering
    \includegraphics[width=16cm]{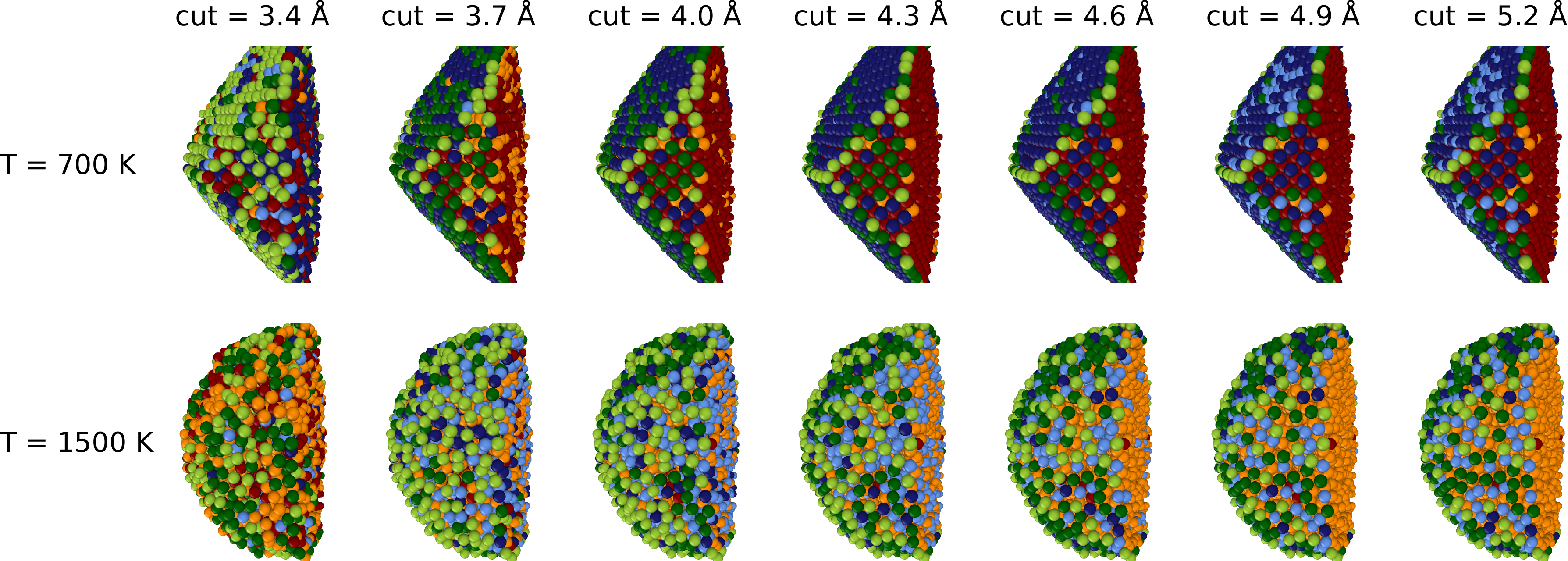}
    \caption{
    \textbf{Atoms in an Au 6266 NP labeled by clustering algorithms employing different cutoff radiuses.}
    Two snapshots of Au 6266 taken the start (top row, T=700~K) and the end (bottom row, T=1500~K) of a MD simulation carried out using the hybrid ML-FF and coloured according to, left to right, a hierarchical clustering algorithm that employs $r\subt{cut}$ of 3.4, 3.7, 4.0, 4.3, 4.6, 4.9, and 5.2~$\text{\AA}$ for the descriptor.}
    \label{fig:clustering_cutoff}
\end{figure*}
\begin{figure*}[b!]
    \centering
    \includegraphics[width=18cm]{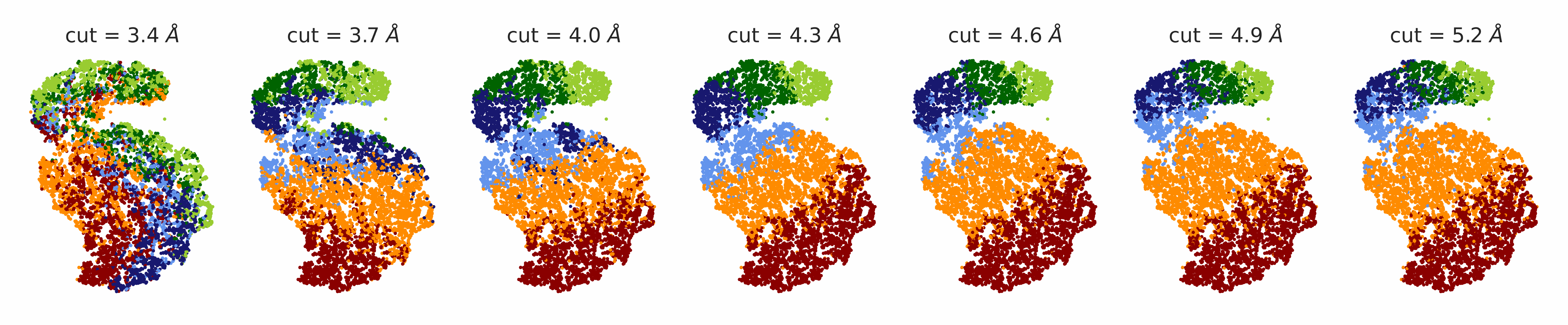}
    \caption{
    \textbf{Consistency of atom labelling across clustering algorithms employing different cutoff radiuses.}
    t-SNE projections mirroring the ones of Figure 1 and Supplementary Figures~\ref{fig:maps_lda} and \ref{fig:maps_hybrid}, for local atomic environments taken from MD simulations carried out using the hybrid ML-FF.
    The x-y coordinates of points are given by the t-SNE projections of 10000 local atomic environment descriptors computed using $r\subt{cut}$ = 4.30 $\text{\AA}$.
    The colours label the six classes assigned by the hierarchical k-means clustering algorithms that employ different $r\subt{cut}$ and mirror the ones employed and defined in the main text.}
    \label{fig:clustering_cutoff_tsne}
\end{figure*}
\begin{figure*}[h!]
    \centering
    \includegraphics[width=17cm]{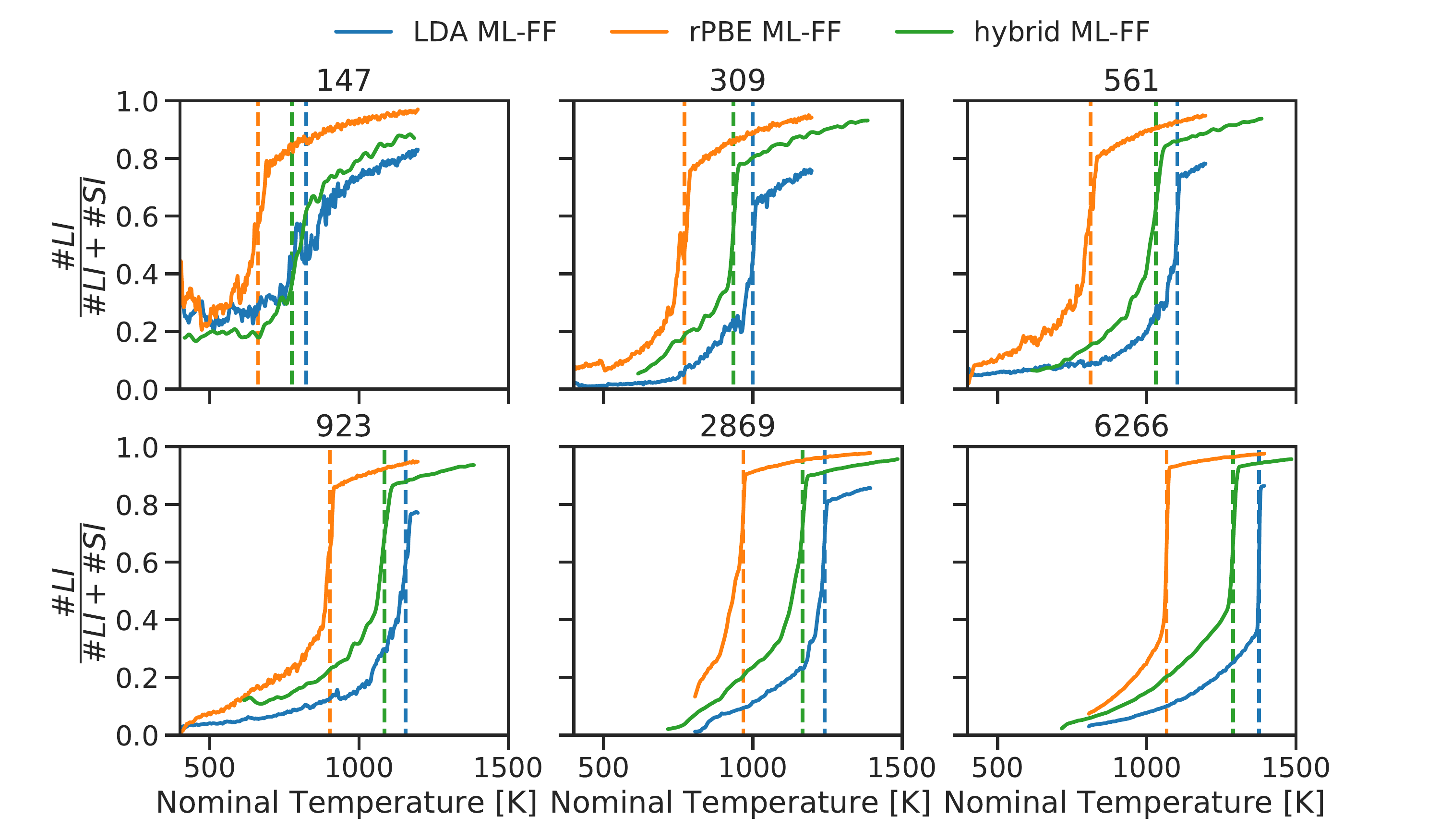}
    \caption{
    \textbf{Evolution of the fraction of liquid inner atoms in MD simulations.}
    Fraction of inner local atomic environment that are classified as liquid by the hierarchical k-means clustering algorithm, as a function of nominal temperature for melting MD simulations carried out using the LDA ML-FF (blue), the rPBE ML-FF (orange), and the hybrid ML-FF (green).
    The vertical dashed lines indicate the $T\subt{melt}\supt{NP}$ obtained with the clustering derivative method.
    All lines are averaged across the repeated simulations.
}
    \label{fig:meltingg}
\end{figure*}
\begin{figure}[!hb]
    \centering
    \includegraphics[width=7.5cm]{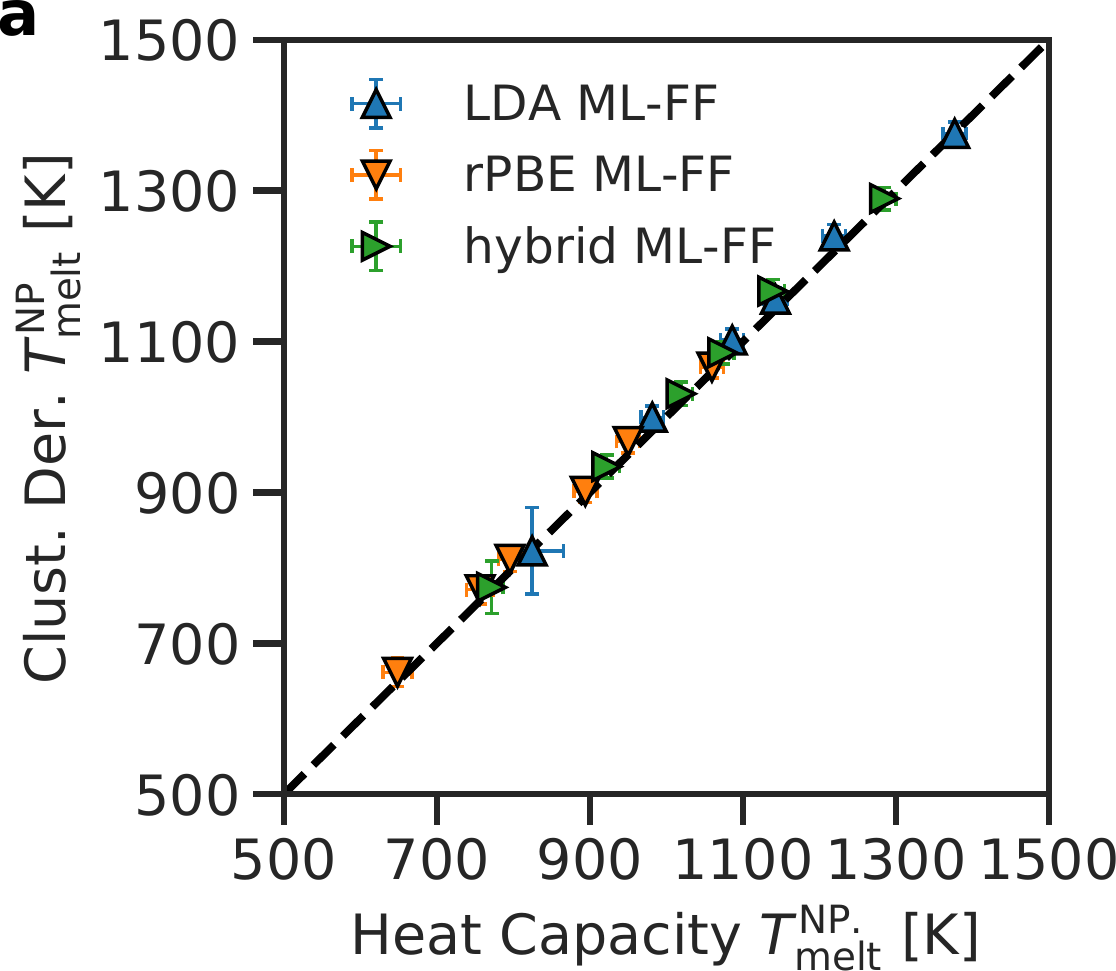}
    \includegraphics[width=7.5cm]{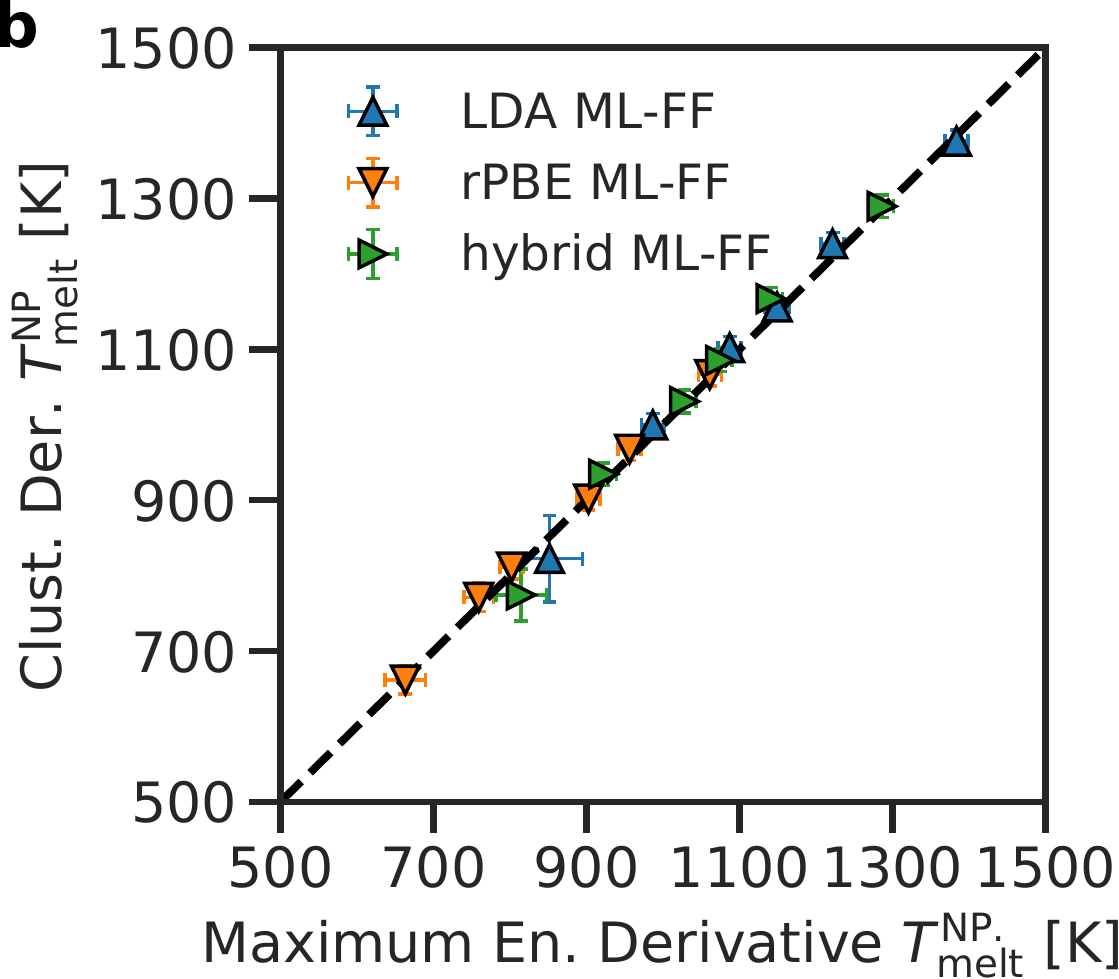}
    \caption{
    \textbf{Comparison of melting temperatures using different estimation methods.}
    Panel a: Pairwise comparison of the $T\subt{melt}\supt{NP}$ for Au NPs computed as the peak of the heat capacity (x-axis) and as the peak in the derivative of the number of inner liquid clusters (y-axis).
    Panel b: Pairwise comparison of the $T\subt{melt}\supt{NP}$ for Au NPs computed as the temperature of the maximum derivative of total energy (x-axis) and as the peak in the derivative of the number of inner liquid clusters (y-axis).
    The error bars show the maximum between the standard deviation of the  $T\subt{melt}\supt{NP}$ estimation across independent MD simulations, and 25~K, the temperature averaging window we employ for the  $T\subt{melt}\supt{NP}$ calculation.
    The black dashed line is a visual aid that indicates a 1:1 correspondence.
}
    \label{fig:melting_t_scatter}
\end{figure}
\begin{figure*}[!htb]
    \centering
    \includegraphics[width=16.2cm]{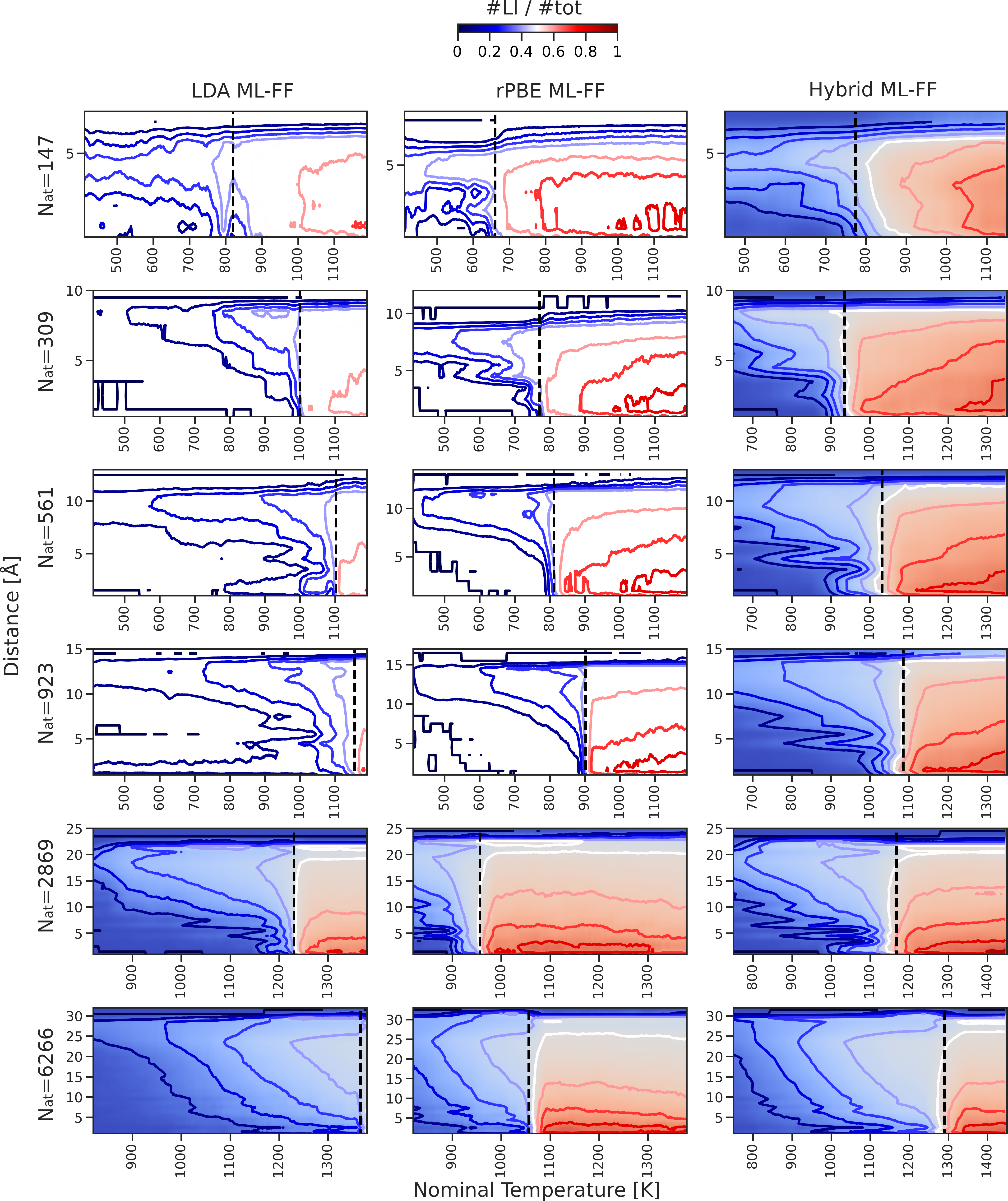}
    \caption{
    \textbf{Evolution of the radial distribution of liquid inner local environments.}
    Average occurrence of the fraction of local atomic environments labeled as LI as a function of radial distance from the center of mass and of nominal MD simulation temperature.
    Plots are shown for the 6 Au NP sizes (increasing from top to bottom), and for the three ML-FFs (from left to right).}
    \label{fig:cluster_heatmap_core}
\end{figure*}
\begin{figure*}[!htb]
    \centering
    \includegraphics[width=16.2cm]{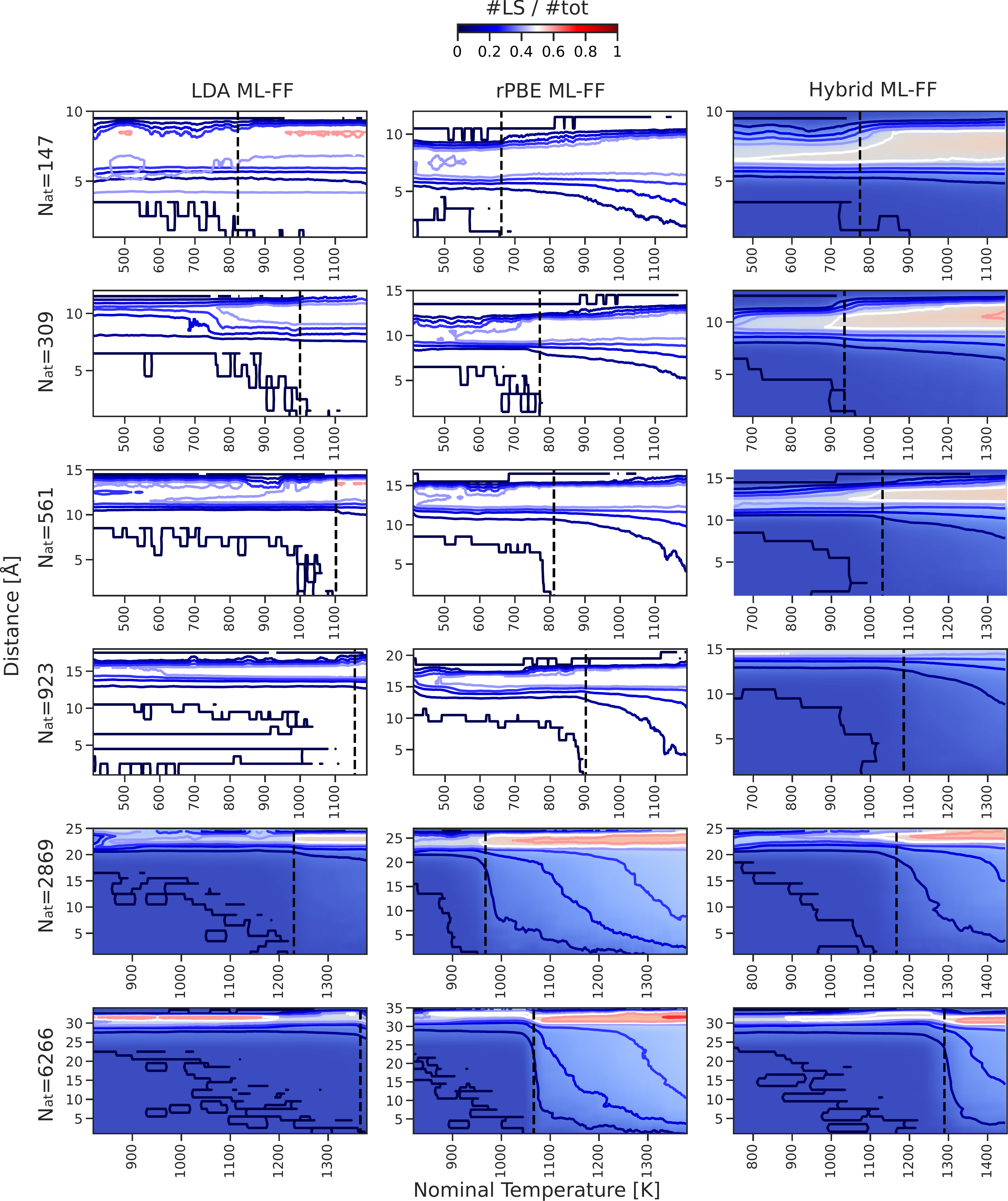}
    \caption{
    \textbf{Evolution of the radial distribution of liquid surface local environments.}
    Average occurrence of the fraction of local atomic environments labeled as LS as a function of radial distance from the center of mass and of nominal MD simulation temperature.
    Plots are shown for the 6 Au NP sizes (increasing from top to bottom), and for the three ML-FFs (from left to right).}
    \label{fig:cluster_heatmap_surf}
\end{figure*}
\begin{figure*}[!htb]
    \centering
    \includegraphics[width=14cm]{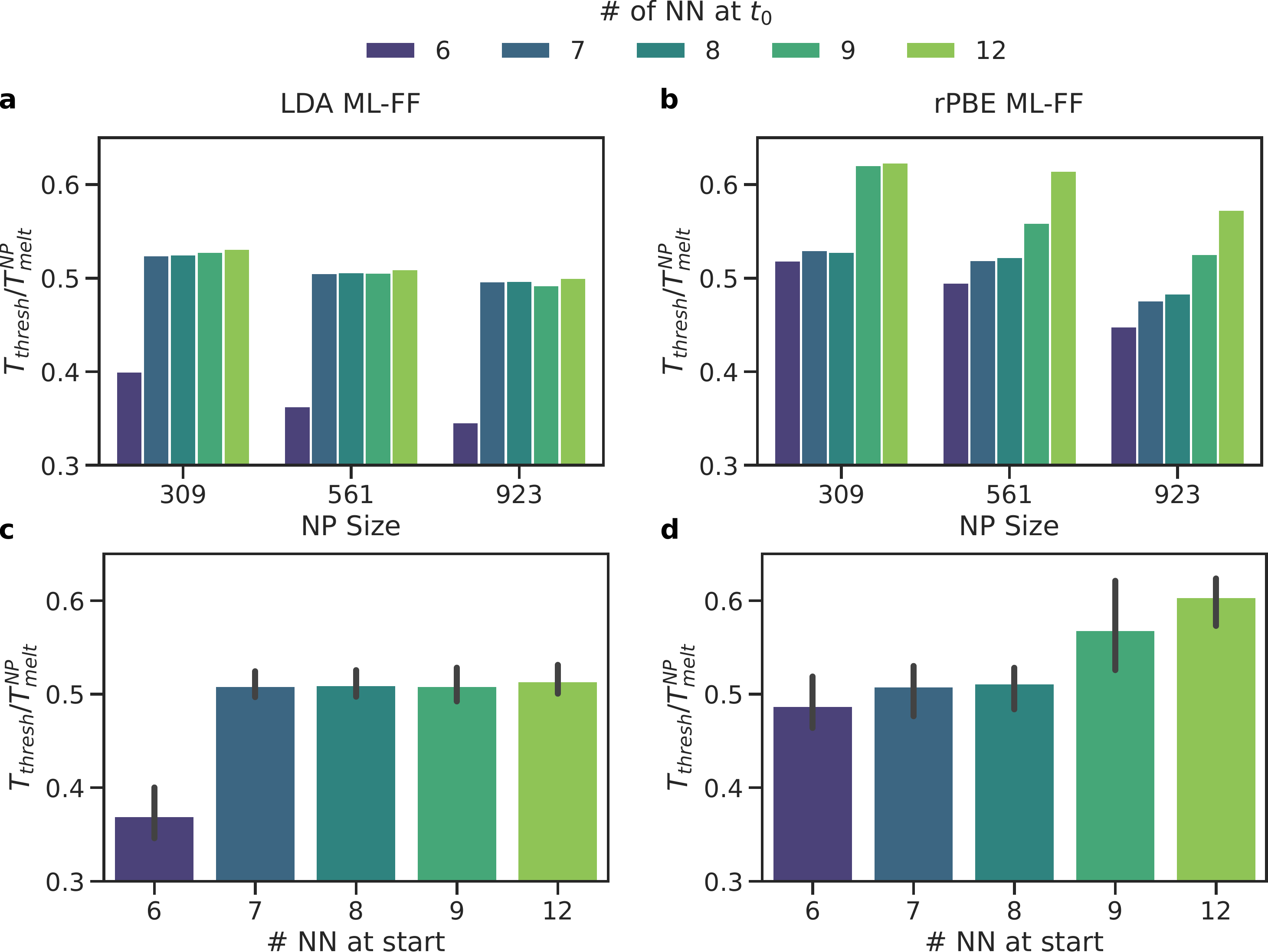}
    \caption{
    \textbf{Effect of initial coordination number on melting kinetics.}
    Ratio between the lowest temperature where \#L / \#tot = 0.4 ($T\subt{thresh}$), and $T\subt{melt}\supt{NP}$ as a function of the initial number of nearest neighbours (\# NN).
    Plots refer to MD simulations of Au 309, 561, and 923 carried out using the LDA ML-FF (panels a and c), and the rPBE ML-FF (panels b and d).
    Panels a and b show the behaviour of $T\subt{thresh}$/$T\subt{melt}\supt{NP}$ averaged over all MD simulations for each size considered, and grouping atoms according to their initial \# NN.
    Panels c and d report the average and the standard deviation of the quantities of panels a and b over the three Au NP sizes considered.}
    \label{fig:t_30_liquid}
\end{figure*}
\begin{figure*}[h!]
    \centering
    \includegraphics[width=17cm]{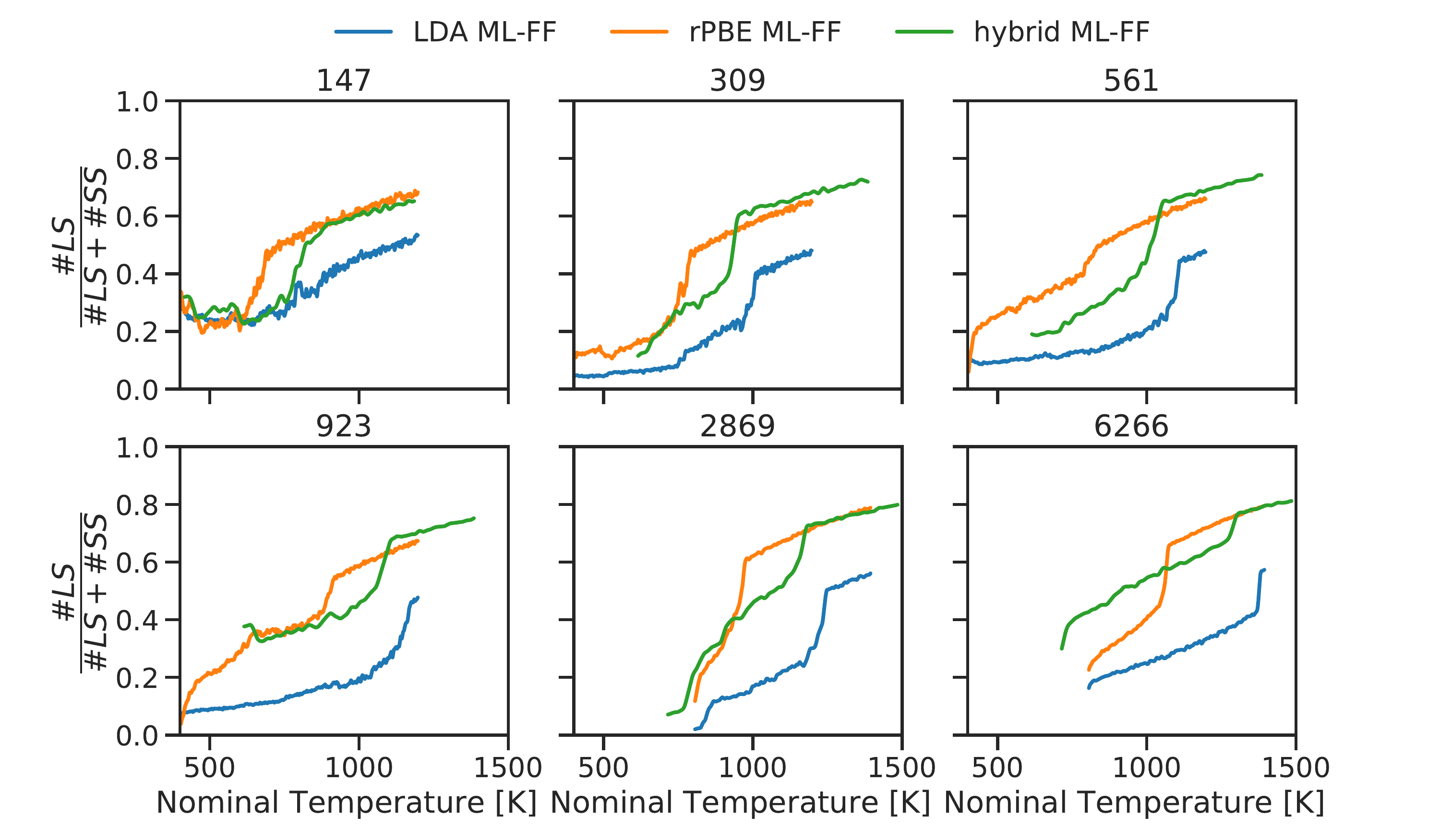}
    \caption{
    \textbf{Evolution of the fraction of liquid surface atoms in MD simulations.}
    Fraction of surface local atomic environment that are classified as liquid by the hierarchical k-means clustering algorithm, as a function of nominal temperature for melting MD simulations carried out using the LDA ML-FF (blue), the rPBE ML-FF (orange), and the hybrid ML-FF (green).}
    \label{fig:melting_surf_evolution}
\end{figure*}

\newpage
\clearpage

\section*{Supplementary Tables}
\begin{table}[h!]
    \centering
    \begin{tabular}{c|c|c|c|c}
    ML-FF & Size & Shape & Force MAE [eV/$\text{\AA}$] & Energy MAE [meV/atom]  \\
    \hline
LDA & 146 &  Oh & 0.10 $\pm$ 0.05 & 7.76  $\pm$ 4.90 \\
 & 147 &  Co &	0.12 $\pm$  0.06    & 5.58  $\pm$ 6.00 \\
 & 147 &  Ih &	0.10 $\pm$  0.05    & 10.79 $\pm$ 5.29 \\
 & 192 & MDh & 0.10 $\pm$  0.05     & 10.93 $\pm$ 5.52 \\
 & 201 &  To &	0.09 $\pm$  0.05    & 9.64  $\pm$ 5.40 \\
 & \textbf{309} &  \textbf{Co} &	\textbf{0.09} $\boldsymbol{\pm}$  \textbf{0.05}    & \textbf{2.65}  $\boldsymbol{\pm}$ \textbf{2.02} \\
 & 561 &  Co & 0.12 $\pm$  0.05     & 2.35  $\pm$ 2.08 \\
\hline
rPBE & 146 &  Oh &	0.09 $\pm$ 0.04 & 6.36 $\pm$ 3.79  \\
 & 147 &  Co &	0.10 $\pm$ 0.05     & 1.91 $\pm$ 1.78  \\
 & 147 &  Ih &	0.08 $\pm$ 0.04     & 8.40 $\pm$ 4.21  \\
 & 192 & MDh & 0.08  $\pm$ 0.04     & 8.59 $\pm$ 4.10  \\
 & 201 &  To &	0.07 $\pm$ 0.04     & 7.28 $\pm$ 4.09  \\
 & \textbf{309} &  \textbf{Co}  &	\textbf{0.07} $\boldsymbol{\pm}$ \textbf{0.03}  & \textbf{1.98} $\boldsymbol{\pm}$ \textbf{1.76}  \\
 & 561 &  Co &	0.09 $\pm$ 0.04     & 2.31 $\pm$ 1.88  \\
    \end{tabular}
    \caption{
    \textbf{Validation errors on forces and energies for the three ML-FFs.}
    Resume of the accuracy (Force and Energy MAE) of ML-FFs trained  on LDA and GGA-rPBE data, for validation sets comprising Au NPs of different sizes and shapes.
    We also report the standard deviation of the MAE on the same validation sets.
    The bold text highlights the size and shape database used to train the ML-FFs.
    Oh: Octahedral, Co: Cube Octahedral, To: Truncated Octahedral, Ih: Icosahedral, MDh: Marks Decahedral.}
    \label{tab:validation}
\end{table}
\newpage
\clearpage

\section*{Supplementary Methods}
\subsection*{Training Database}
In Supplementary Figure~\ref{fig:db}a we show a scatter plot of the first two components yielded by a principal component analysis (PCA) of the 18540 local atomic environments present in the 60 snapshots taken from a MD melting simulation of Au 309 that used to generate the training set of the LDA ML-FFs.
The local atomic environments are described using the 2+3-body atomic cluster expansion representation that employs Bessel functions for the radial basis \cite{Drautz2019, Zeni2021}.
Supplementary Figure~\ref{fig:db}b shows the location on the same projection of 10000 randomly sampled local atomic environments observed during the extensive MD simulations we conducted in NPs of different sizes and for different temperatures employing the LDA ML-FF.
We notice that the training set well covers the space spanned by the local atomic environments sampled from MD simulations and that it contains local atomic environments that belong to all six of the classes identified by the hierarchical k-means clustering  algorithm.

\subsection*{Machine Learning Force Fields Training Curves}
In Supplementary Figure~\ref{fig:learning_curves}, we report the dependency of the Mean Absolute Error (MAE) incurred by the LDA and rPBE ML-FFs trained on an increasing number of de-correlated configurations randomly extracted from an AIMD trajectory, and tested on other configurations sampled during the same MD run, and not employed for training.
The test set contain approximately 15000 local atomic configurations (and forces), and approximately 50 frames (and total energies).
Three independent ML-FFs were trained for each training set size, and in Supplementary Figure~\ref{fig:learning_curves} we display the mean MAEs and their standard deviation across these three repetitions.
We observe that the MAE on force components for both ML-FFs (Supplementary Figure~\ref{fig:learning_curves}a) plateaus around $10^3$ training local atomic environments, corresponding to 3 Au 309 structures.
The MAEs on energy differences remain confined between 2 and 4 meV/atom for the training size range, and present larger fluctuations than the force MAEs.
This is expected, as only one energy reference is associated with each structure, while the force references are 309 (3-components each) per structure (one for each atom in the frame).
Overall, the MAEs incurred for forces and energy differences present negligible variations above $2\cdot10^3$ training points; this justifies our choice of employing 7 Au 309 frames (2163 local atomic environments) for training the ML-FFs.
The convergence of force and energy difference predictions for a rather contained number of data points in the training set is consistent with previous reports utilizing the same framework. \cite{Glielmo2018,Zeni2018}

To further showcase the small dependency of the 3-body ML-FFs on the number of training local atomic environments, we display the 2-body dimer energies in Supplementary Figure~\ref{fig:dimer} for ML-FFs trained on different numbers of frames for the LDA (panel a) and rPBE (panel b) training sets.
The stiffness/softness of the low energy minimum remains essentially unchanged when increasing the number of training points.

\subsection*{Volume-Energy Curves}
In Supplementary Figure~\ref{fig:volume_energy_curves}, we report the cell size-energy curves computed using the three ML-FFs, and the LDA and rPBE-GGA DFT methods, for a FCC lattice containing 32 Au atoms in periodic boundary conditions.
We notice that the equilibrium cell length at zero pressure for the LDA ML-FF (4.035 $\text{\AA}$) nearly matches the experimental value (4.065 $\text{\AA}$), while the rPBE ML-FF has a larger equilibrium cell length (4.225 $\text{\AA}$).
Moreover, we notice how the LDA ML-FF is stiffer and more bound than the rPBE ML-FF; this effect can also be noticed by visualizing the 2+3-body ML-FF directly as seen in Supplementary Figure~\ref{fig:pes}.

The ML predictions are in qualitative agreement with the DFT ground truth data.
From a quantitative standpoint, the ML-FFs are found to underestimate (i.e. predict more negative) cohesive energies, with respect to both rPBE and LDA training sets.
We nevertheless note that the ML-FFs' accuracies are acceptable on the grounds that no data of periodic systems were utilized to train the ML models.
The hybrid ML-FF presents equilibrium cell energy at zero pressure that matches the experimental value; this is expected, as the hybrid ML-FF was fitted to match this experimental property.

We also calculate the bulk modulus of FCC Au at 0~K for the three ML-FFs as \cite{meyers2008mechanical}:
\begin{equation}
K = \Omega_0 \left( \dfrac{\partial \varepsilon}{\partial \Omega^2} \right)_{\Omega_0},
\end{equation}
where $\Omega$ is the atomic volume, and the subscript 0 indicates the equilibrium atomic volume at 0~K.
We obtain values of 203, 170, and 153 GPa for the bulk moduli yielded by the LDA, rPBE and hybrid ML-FFs, respectively, whereas the experimental value for FCC Au ranges from 142 to 220 GPa, depending on the source. \cite{kittel2005crystal, kelly2014properties, samsonov2012handbook}

\subsection*{Machine Learning Potential Energy Surfaces}
Mappable (2+3)-body ML-FFs are inherently interpretable since they are equivalent to a 2+3-body tabulated potential.
From the latter, it is trivial to probe whether the interatomic (2+3)-body interactions result, e.g., in a stiff or in a soft potential.
In Supplementary Figure~\ref{fig:pes}, we report the 2D distance map for the three ML-FFs here developed, where we notice how the LDA ML-FF has a stiff and strong interaction between atoms, while the rPBE ML-FF shows a shallower potential energy surface, which also results in weaker bond strength.
The hybrid ML-FF has, as expected, a shape that is in-between the one of the LDA and rPBE ML-FF.
These characteristics are reflected in the $T\subt{melt}\supt{NP}$ we estimate for Au NPs, as the LDA ML-FF always predicts higher $T\subt{melt}\supt{NP}$ values than the hybrid ML-FF, which in turn predicts higher $T\subt{melt}\supt{NP}$ values than the rPBE ML-FF.

\subsection*{Machine Learning Force Fields Validation}
In Supplementary Table \ref{tab:validation}, we report the validation MAE on energy differences and force components incurred by the LDA-trained and r-PBE-trained ML-FFs on datasets gathered from classical molecular dynamics (MD) trajectories previously discussed in \citet{Delgado-Callico2020} and in \citet{Foster2019}, where nanoparticles of different sizes (leftmost column) and initial morphology (second column from the left) undergo solid-solid and solid-liquid transitions.
For each NP size and each morphology, a variable number of structures has been selected at random; the total number of local atomic environments used in validation is approximately 5000 for each NP size and shape.

\subsection*{Effect of Heating Rate on Melting Temperature}
All of the results and data we show refer to melting MD simulations where the heating rate was kept constant at 20~K/ns.
Here, we briefly discuss the effect of heating rate on the $T\subt{melt}\supt{NP}$ for Au NPs containing 147, 309, 561, 923 and 6266 atoms due to super-heating.
In Supplementary Figure~\ref{fig:heating_rate}, we report the $T\subt{melt}\supt{NP}$ computed for MD simulations of Au 147, 309, 561, 923, and 6266 carried out using the LDA ML-FF with a heating rate of 20~K/ns, 10~K/ns for all sizes, and also 5~K/ns for NPs with less than 923 atoms.
We observe that the heating rate has little effect on the $T\subt{melt}\supt{NP}$; this, therefore, reinforces our belief the super-heating effects are not strongly affecting our $T\subt{melt}\supt{NP}$ estimates.

\subsection*{Clustering of Local Atomic Environments}
Supplementary Figure~\ref{fig:pca1} shows five 2-dimensional PCA projections of 10000 local atomic environments randomly sampled from the MD simulations of Au 147, 309, 561, 923, 2869, 6266 carried out using the rPBE ML-FF.
Local atomic environments are described using the 2+3-body atomic cluster expansion representation that employs Bessel functions for the radial basis set \cite{Drautz2019, Zeni2021}.
More specifically, we expand the local atomic environment density into 4 radial bases and 4 angular bases and employ a radial cutoff of 4.24 (4.42, 4.30) $\text{\AA}$ for LDA (r-PBE, hybrid) ML-FF MD simulation data, resulting in a 40-dimensional representation.
In the first map, the colour coding shows the label obtained by utilizing a k-means clustering with two clusters (k=2) on the full set of local environments; at this step inner from surface local atomic environments are discriminated.
At the second iteration of k-means clustering, applied only to local atomic environments labelled as surface by the previous clustering step, a division between high- and low-coordinated surface environments emerges. 
When the final clustering iteration is carried out, solid and liquid environments in each group are discriminated against.

\subsection*{Clustering for LDA-trained ML-FF}
In Supplementary Figure~\ref{fig:maps_lda} we show the parallel to Figure 3 in the main text, for the case of local atomic environments sampled from MD simulations carried out using the LDA-trained ML-FF. 
Also in this case a clear separation of the six families of local environments arises.

\subsection*{Clustering for hybrid ML-FF}
In Supplementary Figure~\ref{fig:maps_hybrid}, we show the parallel to Figure 3 in the main text, for the case of local atomic environments sampled from MD simulations, carried out using the hybrid ML-FF. 
Also in this case a clear separation of the six families of local environments arises.

\subsection*{Effect of cutoff radius on Clustering}

To assess the sensitivity of our local atomic environment classification algorithm to the choice of the cutoff radius of the descriptor, we perform hierarchical clustering on data coming from MD simulations carried out with the hybrid ML-FF, and using cutoff radius ($r\subt{cut}$) values ranging from 3.4~$\text{\AA}$ to 5.2~$\text{\AA}$, every 0.3~$\text{\AA}$.
In Supplementary Figure~\ref{fig:clustering_cutoff}, we display two snapshots taken from the start (top row, T=700~K) and the end (bottom row, T=1500~K) of a MD simulation of Au 6266 carried out using the hybrid ML-FF, where the atoms are coloured according to hierarchical clustering schemes that employ increasing $r\subt{cut}$, from left to right. 
We notice that the classification is coherent for $r\subt{cut} >$ 4.0 $\text{\AA}$ in this example.
This is somehow expected, as information about the position of second neighbours is key in identifying the melting (see manuscript and  \cite{Delgado-Callico2020}).
In Supplementary Figure~\ref{fig:clustering_cutoff_tsne}, we display the t-SNE plot of the local atomic environment descriptor using $r\subt{cut}$ = 4.30 $\text{\AA}$ and labelled using hierarchical k-means clustering algorithms that employ different $r\subt{cut}$ values, for atoms taken from MD simulations carried out using the hybrid ML-FF, coloured according to the class assigned by the clustering algorithm.
Also in this case, the labeling is coherent for $r\subt{cut} > $ 4.0.

\subsection*{Melting Temperature Estimate}
In Supplementary Figure~\ref{fig:meltingg}, we report the temporal evolution of the fraction of inner local atomic environments that are classified as liquid, averaged over all MD simulations carried out for each NP size and each ML-FF.
A triangular rolling average of width 50~K was used to smooth out the data to ease the visualization, and to make the  $T\subt{melt}\supt{NP}$ estimate more robust.
This smoothing introduces and uncertainty, which we consider equal to the standard deviation of the triangular distribution used, i.e. 12~K.
For all Au NPs except Au147 and both ML-FFs, the value of \#LI / (\#LI+\#SI) has a sharp transition, where the majority of inner local atomic environments becomes liquid.
We can numerically identify such transition, which we claim is indeed the $T\subt{melt}\supt{NP}$, as the temperature where the maximum positive derivative of \#LI / (\#LI+\#SI)  w.r.t. the nominal simulation temperature (or, equivalently, simulation time) is observed.
To numerically estimate such temperature, we revert to finite differences to avoid the complications in taking derivatives of noisy data.
In particular, we estimate the derivative of \#LI / (\#LI+\#SI) at temperature $T^*$ as the difference between its value at temperature $T^* + 25~K$ and $T^* - 25~K$, divided by 50~K.
We refer to this method to estimate the $T\subt{melt}\supt{NP}$ as the clustering derivative method.
In Supplementary Figure~\ref{fig:meltingg} we highlight the $T\subt{melt}\supt{NP}$ estimated using this approach using vertical dashed lines, with colours matching the ones of the \#LI / (\#LI+\#SI) lines and identifying the different MD simulations.

In Supplementary Figure~\ref{fig:melting_t_scatter}, we report the pairwise correspondences between the $T\subt{melt}\supt{NP}$ computed using our clustering derivative method, and two commonly used algorithms.
The first, named Heat Capacity in Supplementary Figure~\ref{fig:melting_t_scatter}, estimates the $T\subt{melt}\supt{NP}$ as the temperature where a peak in the heat capacity is observed. \cite{Delgado-Callico2020, chen2020heating}
The second, named Maximum En. Derivative in Supplementary Figure~\ref{fig:melting_t_scatter}, estimates the $T\subt{melt}\supt{NP}$ as the temperature where the highest standard deviation of the total energy is observed. \cite{Delgado-Callico2020, chen2020heating}
The three methods yield $T\subt{melt}\supt{NP}$ predictions that align almost perfectly.

\subsection*{Liquid Environments Evolution}
Supplementary Figures~\ref{fig:cluster_heatmap_core} and Figure~\ref{fig:cluster_heatmap_surf} show the dependence on the nominal simulation temperature (x-axis) and the distance from the centre of mass (y-axis) of the fraction of LI and LS local atomic environments (colour), respectively,  for all NP sizes and the three ML-FFs.

\subsection*{Effect of Initial Coordination Number on Melting Kinetics}
In Supplementary Figure~\ref{fig:t_30_liquid}, we report the lowest temperatures where \#L / \#tot = 0.4 ($T\subt{thresh}$), normalized by the $T\subt{melt}\supt{NP}$ found for each NP by the appropriate ML-FF.
The $T\subt{thresh}$/$T\subt{melt}\supt{NP}$ are computed for atoms that have been divided into groups according to the number of nearest neighbours (\#NN) they possess at the start of each MD simulation ($t_0$).
Panels a and b display the $T\subt{thresh}$/$T\subt{melt}\supt{NP}$ as a function of NP size, while the size-averaged $T\subt{thresh}$/$T\subt{melt}\supt{NP}$ are displayed in panels c and d, for the LDA (a, c) and rPBE (b, d) ML-FFs.
We observe that the relative temperature at which at least 40\% of atoms are labelled as liquid increases with the initial coordination number.
In particular, atoms starting on the edges of the NPs (\#NN=6) become liquid at significantly lower temperatures than the ones starting at the surface (\#NN=7, 8, 9), which in turn reach the \#L / \#tot = 0.4 liquid threshold at temperatures below the ones for inner atoms (\#NN=12).
The mean first-passage temperature to move into a liquid phase is also dependent on the system size, as surface atoms in smaller NPs display, on average, lower MFPTs than the one found for larger NPs.
The present observations hold regardless of the ML-FF utilized. Nonetheless, we notice that an rPBE-based description of the interatomic interactions translates in trajectories where the change into a liquid phase is delayed w.r.t. the case of dynamical evolution sampled via an LDA-based ML-FF.

\subsection*{Surface Phase Change}
In Supplementary Figure~\ref{fig:melting_surf_evolution}, we report the temporal evolution of the fraction of surface local atomic environments that are classified as liquid, averaged over all MD simulations carried out for each NP size and each ML-FF.
A triangular rolling average of width 50~K was used to smooth out the data to ease the visualization.
We observe that, in contrast with the inner local environments of Supplementary Figure~\ref{fig:meltingg}, the lines of Supplementary Figure~\ref{fig:melting_surf_evolution} present small positive jumps, and often at temperatures that match the $T\subt{melt}\supt{NP}$.
This may suggest that no first-order phase transition other than the melting phase transition at $T\subt{melt}\supt{NP}$ takes place in the NPs.
For this reason, it is not advisable to define a surface melting temperature for the systems we simulate, but rather to evaluate the temperatures at which a sizeable fraction of the surface environments is labelled as liquid.
\end{document}